\documentclass[prx,twocolumn,floatfix,superscriptaddress,longbibliography,notitlepage, nofootinbib, aps, 10pt]{revtex4-2}
\usepackage{academicons}
\usepackage{algorithm2e}
\usepackage{graphicx, color, graphpap}
\usepackage{enumitem}
\usepackage{amssymb}
\usepackage{amsthm}
\usepackage{braket}
\usepackage{dsfont}
\usepackage{mathtools}
\usepackage[dvipsnames]{xcolor}
\usepackage{multirow}
\usepackage[colorlinks=true,citecolor=blue,linkcolor=magenta]{hyperref}
\usepackage[T1]{fontenc}
\usepackage{bbm}
\usepackage{thmtools,thm-restate}
\usepackage{verbatim}
\usepackage{mathtools}
\usepackage{titlesec}
\usepackage{amsmath}
\usepackage{subfigure}
\usepackage{orcidlink}
\usepackage{tabularx}
\usepackage{booktabs}
\usepackage[thinlines]{easytable}  
\usepackage{listings}
\usepackage{csquotes}
\makeatletter 
\renewcommand\onecolumngrid{
\do@columngrid{one}{\@ne}
\def\set@footnotewidth{\onecolumngrid}
\def\footnoterule{\kern-6pt\hrule width 1.5in\kern6pt}
}

\renewcommand\twocolumngrid{
        \def\footnoterule{
        \dimen@\skip\footins\divide\dimen@\thr@@
        \kern-\dimen@\hrule width.5in\kern\dimen@}
        \do@columngrid{mlt}{\tw@}
}

\newcommand{\RemoveAlgoNumber}{\renewcommand{\fnum@algocf}{\AlCapSty{\AlCapFnt\algorithmcfname}}}
\newcommand{\RevertAlgoNumber}{\algocf@resetfnum}

\makeatother
\newcommand{\norm}[1]{\left| \left| #1 \right| \right|}
\newcommand{\comm}[2]{\left[ #1 , #2 \right]}
\newcommand{\Tr}{{\rm Tr}}
\newcommand{\abs}[2][]{#1| #2 #1|}

\newtheorem{lemma}{Lemma}

\hypersetup{
pdftitle={Large-scale simulations of Floquet physics on near-term quantum computers},
pdfsubject={Quantum Computing},
pdfauthor={Timo Eckstein, Refik Mansuroglu, Piotr Czarnik, Jian-Xin Zhu, 
Michael J. Hartmann, Lukasz Cincio, Andrew T. Sornborger, and Zo\"{e} Holmes},
pdfkeywords={Quantum computing, NISQ, Hamiltonian simulation, Quantum algorithm, Floquet physics, Many-body physics}
}
\begin{document}
\title{Large-scale simulations of Floquet physics on near-term quantum computers} 

\author{Timo Eckstein 
\orcidlink{0000-0002-6819-1865}}
\email{Timo.Eckstein@fau.de}
\affiliation{Department of Physics, Friedrich-Alexander Universität Erlangen-Nürnberg, Erlangen, Germany}
\affiliation{Max-Planck Institute for the Science of Light, Erlangen, Germany}
\affiliation{Theoretical Division, Los Alamos National Laboratory, Los Alamos, New Mexico 87545, USA}

\author{Refik Mansuroglu
\orcidlink{0000-0001-7352-513X}}
\affiliation{Department of Physics, Friedrich-Alexander Universität Erlangen-Nürnberg, Erlangen, Germany}

\author{Piotr Czarnik
\orcidlink{0000-0002-0477-1158}}
\affiliation{Faculty of Physics, Astronomy, and Applied Computer Science, Jagiellonian University, Krak\'ow, Poland}
\affiliation{Mark Kac Center for Complex Systems Research, Jagiellonian University, Krak\'ow, Poland}

\author{Jian-Xin Zhu
 \orcidlink{0000-0001-7991-3918}}
\affiliation{Theoretical Division, Los Alamos National Laboratory, Los Alamos, New Mexico 87545, USA}
\affiliation{Center for Integrated Nanotechnologies, Los Alamos National Laboratory, Los Alamos, New Mexico 87545, USA}

\author{Michael J. Hartmann
\orcidlink{0000-0002-8207-3806}}
\affiliation{Department of Physics, Friedrich-Alexander Universität Erlangen-Nürnberg, Erlangen, Germany}
\affiliation{Max-Planck Institute for the Science of Light, Erlangen, Germany}

\author{Lukasz Cincio
\orcidlink{0000-0002-6758-4376}}
\affiliation{Theoretical Division, Los Alamos National Laboratory, Los Alamos, New Mexico 87545, USA}

\author{Andrew T. Sornborger
 \orcidlink{0000-0001-8036-6624}}
\affiliation{Information Sciences, Los Alamos National Laboratory, Los Alamos, New Mexico 87545, USA}

\author{Zo\"{e} Holmes
\orcidlink{0000-0001-6841-4507}}
\affiliation{Theoretical Division, Los Alamos National Laboratory, Los Alamos, New Mexico 87545, USA}
\affiliation{Laboratory of Quantum Information and Computation,  École Polytechnique Fédérale de Lausanne, Lausanne, Switzerland}

\begin{abstract}
Periodically driven quantum systems exhibit a diverse set of phenomena but are more challenging to simulate than their equilibrium counterparts. Here, we introduce the Quantum High-Frequency Floquet Simulation (QHiFFS) algorithm as a method to simulate fast-driven quantum systems on quantum hardware. Central to QHiFFS is the concept of a kick operator which transforms the system into a basis where the dynamics is governed by a time-independent effective Hamiltonian. This allows prior methods for time-independent simulation to be lifted to simulate Floquet systems. We use the periodically driven biaxial next-nearest neighbor Ising (BNNNI) model, a natural test bed for quantum frustrated magnetism and criticality, as a case study to illustrate our algorithm. We implemented a 20-qubit simulation of the driven two-dimensional BNNNI model on Quantinuum’s trapped ion quantum computer. Our error analysis shows that QHiFFS exhibits not only a cubic advantage in driving frequency $\omega$ but also a linear advantage in simulation time $t$ compared to~Trotterization.
\end{abstract}

\maketitle
 
\section{Introduction}
\begin{figure}[t!]
    \centering
    \includegraphics[width=\linewidth]{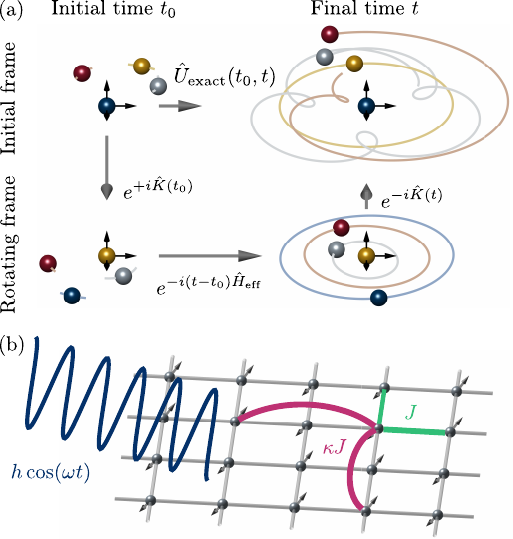}
    \caption{\textbf{Content overview.}
    (a) QHiFFS Algorithm Concept. To realize a time-dependent periodic time evolution, we transform from the standard basis to a (periodically) rotating frame using $e^{+i\hat{K}(t_0)}$. Evolution in this frame from $t_0$ to $t$ is governed by the time-independent effective Hamiltonian $\hat H_\mathrm{eff}$. The adjoint of the kick operator $e^{-i\hat{K}(t)}$ is applied at the final simulation time $t$ to kick back into the original frame. 
    (b)~Benchmark system. To demonstrate the advantage of QHiFFS, as compared to standard Trotterization, we implement the 20-qubit transversely driven 2D biaxial next-nearest neighbor Ising (BNNNI) model with periodic boundary conditions on Quantinuum's H1 trapped ion quantum hardware (see Fig.~\ref{fig: results}). The BNNNI model, Eq.~\eqref{eq:BNNNI_Hamiltonian}, consists of nearest-neighbor coupling with strength $J$ and next-nearest-neighbor coupling with strength $\kappa J$. We suppose a 4 by 5 qubit lattice is periodically driven with strength $h$ and at frequency $\omega$.}
    \label{fig:ansatz_idea}
\end{figure}

There has been a recent spike of interest in the dynamical control of quantum phases by driving strongly correlated quantum systems out of equilibrium. 
In particular, Floquet systems, that is quantum systems subjected to periodic driving, not only exhibit a rich set of physical phenomena, including exotic non-equilibrium topological phases of matter~\cite{khemani2016phase,moessner2017equilibration, moessner2021topological}, but can also generate quantum phenomena on demand for technological applications~\cite{lindner2011floquet,mentink2015ultrafast,mitrano2016possible,basov2017towards,mciver2020light}.
For example, time crystalline phases~\cite{wilczek2012quantum}, which break continuous time-translational symmetry have been demonstrated with trapped ions~\cite{zhang2017observation}, superconducting qubits~\cite{ippoliti2021many, mi2022time, frey2022realization} and diamond defects~\cite{randall2021many}. 
Similarly, periodic driving can lead to symmetry-protected edge modes~\cite{khemani2016phase, moessner2017equilibration, moessner2021topological}, a phenomenon that has been realized in both trapped ions~\cite{dumitrescu2022dynamical} and superconducting~\cite{zhang2022digital} platforms, and has potential applications for error correction. 
However, the strong correlations present in such systems make their modeling classically challenging. 

While analog quantum simulators have been employed to study non-equilibrium dynamics \cite{jotzu2014experimental, struck2011quantum,weitenberg2021tailoring}, they suffer from an intrinsic model-dependency. Digital quantum simulation promises universal algorithms for non-equilibrium dynamics with exponentially less resources than known classical methods. In fact, this provided one of the original motivations for developing quantum hardware in the first place~\cite{feynman1982simulating}. While a distant dream for decades, with rapid technological progress, the possibility of realizing a quantum advantage for quantum simulation is now in sight~\cite{arute2019quantum, zhong2020quantum}. Nonetheless, error rates are yet to reach the error correction threshold for moderate-sized devices and hence quantum simulation algorithms designed for fault-tolerant devices remain challenging for the near future~\cite{lloyd1996universal, sornborger1999higher, verstraete2009quantum, berry2015simulating}. This has prompted the search for quantum simulation algorithms suitable for implementation on near-term quantum devices~\cite{verstraete2009quantum, cirstoiu2020variational, gibbs2021long, gibbs2024dynamical, mansuroglu2023variational, barison2021efficient}. However, most of these studies focused on simulating time-independent systems and less attention has so far been paid to how to simulate periodically driven systems.

Standard Trotterization~\cite{khemani2016phase, moessner2017equilibration, moessner2021topological, dumitrescu2022dynamical, zhang2022digital, wilczek2012quantum, zhang2017observation, ippoliti2021many, mi2022time, frey2022realization, randall2021many, schneider2012quantum, fauseweh2021digital, lamm2018simulation, oftelie2020towards, rodriguez2022real}, where the time-ordered integral governing the dynamics is discretized into short time-steps (implemented via a Trotter approximation), fails to take advantage of the periodic structure of Floquet problems and thus acquires a substantial computational overhead.

Alternative proposals include qubitization of the truncated Floquet-Hilbert space~\cite{fauseweh2023quantum}, which require additional ancillae, truncated Taylor~\cite{berry2015simulating} and Dyson series~\cite{low2019hamiltonian} which are not guaranteed to be unitary or using variational optimization~\cite{lamm2018simulation}. The former require deep circuits, typically not accessible in the NISQ era and the latter is limited by (noise-induced) barren plateaus~\cite{mcclean2018barren,cerezo2020cost,holmes2021connecting, holmes2021barren, arrasmith2021equivalence, wang2020noise} as well as other barriers to variational optimization on near-term quantum hardware~\cite{Bittel2021Training, anschuetz2022quantum}. Correspondingly, digital quantum simulations of systems subject to a continuous driving field have thus far been limited to only a few qubits~\cite{oftelie2020towards, rodriguez2022real}.

Here we draw motivation from prior work on classical methods for studying Floquet physics~\cite{floquet1883quations, magnus1954exponential, shirley1965solution, casas2001floquet, goldman2014periodically, eckardt2015high, oka2019floquet} and methods for analog quantum simulation~\cite{schneider2012quantum}, to develop a quantum algorithm for simulating periodically driven systems on digital quantum computers.
 Crucially this approach takes explicit advantage of the periodicity of the driving terms to reduce the resource requirements of the simulation. Key to our approach, is the use of a high-frequency approximation that performs a time-dependent basis transformation into a rotating frame~\cite{goldman2014periodically, eckardt2015high}. In effect we reverse the usual perspective taken by Floquet engineering, where analog simulation of time-dependent Hamiltonians is used to implement the effective time-independent dynamics. Here we instead take advantage of this correspondence to implement the effective time-independent dynamics on quantum hardware using a kick operator. Previously developed time-independent techniques can then be lifted and reused for the Floquet simulation~\cite{lloyd1996universal, sornborger1999higher,verstraete2009quantum, berry2015simulating,cirstoiu2020variational, gibbs2021long, mansuroglu2023variational, gibbs2024dynamical}. The Quantum High Frequency Floquet Simulation (QHiFFS – pronounced \textit{quiffs}) algorithm thus opens up the potential of simulating fast-driven systems using substantially shorter-depth circuits than standard Trotterization. 

We use QHiFFS to implement a 20-qubit, 230-gate two-dimensional simulation of the transversely driven biaxial next-nearest neighbor Ising (BNNNI) model on a trapped ion quantum computer. This implementation is, to the best of our knowledge, the largest digital 2D quantum simulation of a periodically driven system to date (see Sec.~\ref{sec: results}). We choose to focus on the BNNNI model because it is one of the most representative spin-exchange models containing both tunable frustration and tunable quantum fluctuations, creating competing quantum phases and criticality. Indeed, this behavior is successfully captured by our hardware implementation. We stress that this simulation would not have been feasible using standard Trotterization, which would have required a 50-fold deeper circuit (see Fig.~\ref{fig: results}).  

Finally, we set QHiFFS on solid analytic foundations by providing an analysis on the scaling of the final simulation errors. In particular, we find that first-order QHiFFS not only exhibits a cubic scaling advantage in driving frequency $\omega$, but also a linear one in simulation time $t$ compared to a second-order Trotter sequence. The error scaling in system size $n$ remains linear for local models (see Error Analysis). Thus we see QHiFFS finding use both on moderate-sized error-prone near-term devices and in the fault-tolerant era. 

\section{Results}
\subsection{The QHiFFS Algorithm.} 
\label{sec:kick_ansatz}
\label{sec: algorithm}
The QHiFFS algorithm simulates quantum systems governed by a Hamiltonian of the form 
\begin{align}
\hat H(t) = \hat H_0 + \hat V(t) \quad \mathrm{with} \quad \hat V(T+t) = \hat V(t) \, ,
\label{eq: Floquet_Hamiltonian}
\end{align}
where $\hat H_0$ is the time independent non-driven Hamiltonian and $\hat V(T+t)$ is a periodic driving term. Our aim is to simulate real-time evolution under this Hamiltonian from some initial time $t_0$ to some final time $t$. That is, to implement the unitary 
\begin{align}
\hat U_{\rm exact} (t_0, t)   &= \mathcal{T} \exp\left( -i \int_{t_0}^{t} \hat H(s) ds \right) \, ,
\label{eq:time_ordered_integral} 
\end{align} 
where $\mathcal{T}$ is the time ordering operator. 

As is often found in physics, choosing the right reference frame is key to a simple description of the dynamics of a system. To take a paradigmatic example - the motion of planets in our solar system appears rather complex in the earth's reference frame but becomes much simpler if we transform into the frame of reference of the sun. Our algorithm, as sketched in Fig.~\ref{fig:ansatz_idea}, takes a similar approach~\cite{whaley1984rotating}: 
\begin{enumerate}
    \item Apply $e^{+i\hat K(t_0)}$ to \textit{kick} an initial state $|\psi\rangle$ at time $t_0$ into a reference frame where the dynamics of the driven system is governed by an \textit{effective Hamiltonian} $\hat H_{\rm eff}$.
    \item Apply $e^{-i(t - t_0)\hat H_{\rm eff}}$ to evolve the state from times $t_0$ to $t$ in this new reference frame.
    \item Apply $e^{-i\hat K(t)}$ to \textit{kick} the system back into the original (lab) frame of reference. 
\end{enumerate}
Or, more compactly, the total simulation is of the form 
\begin{align}
\hat U_{\rm exact} (t_0, t)  &= e^{-i\hat K(t)}e^{-i(t - t_0)\hat H_{\rm eff}}e^{+i\hat K(t_0)}  \, .
\label{eq: Kick_unitary}
\end{align}
Crucially, $\hat H_{\rm eff}$ is time-independent. Thus, one can use prior methods~\cite{lloyd1996universal, sornborger1999higher,verstraete2009quantum, berry2015simulating,cirstoiu2020variational, gibbs2021long, mansuroglu2023variational, gibbs2024dynamical} for time-independent quantum simulation for its implementation. Furthermore, in many cases, as discussed below, the kick operator, $\hat K(t)$, will take a simple form enabling it to be easily implemented on quantum hardware. 

The kick transformation and corresponding effective Hamiltonian for a periodically-driven system can be obtained by expanding perturbatively in the driving frequency $\omega = 2 \pi/T$~\cite{casas2001floquet, goldman2014periodically, goldman2015erratum,eckardt2015high}. As detailed in Methods and \ref{app:derivation}, to $\mathcal O(\frac{1}{\omega^2})$ the truncated effective Hamiltonian $\tilde H_{\rm eff}$ and kick operator $\tilde K(t)$ are given by:
\begin{align}
\tilde H_{\rm eff} &\approx \hat H_0 + \frac{1}{\omega} \sum_{j=1}^\infty \frac{1}{j} [\hat V^{(j)},\hat V^{(-j)}]  
     \label{eq: Heff_series_1st_order}\\
\tilde K(t) &\approx -\frac{i}{\omega}\sum_{j=1}^\infty \frac{1}{j} \left( \hat V^{(j)}e^{ij\omega t}  - \hat V^{(-j)}e^{-ij\omega t} \right)   \, .
\label{eq: Kick_series_1st_order}
\end{align}
with Fourier components $\hat V^{(j)}$ of the time-dependent potential $\hat V(t)  = \sum_{j=1}^\infty \hat V^{(j)} e^{ij\omega t} +  \hat V^{(-j)} e^{-ij\omega t}$. 
We note that the rotating wave approximation takes an analogous approach in the sense that under the rotating wave approximation evolution is governed by the $0^{\rm th}$ order approximation to the propagator in the interaction picture.

For analog quantum simulators, where Floquet engineering is commonly applied, introducing particular periodic drivings on native analog physical systems~\cite{struck2011quantum, jotzu2014experimental, weitenberg2021tailoring} is relatively easy and so Eq.~\eqref{eq:time_ordered_integral} is used to approximate Eq.~\eqref{eq: Kick_unitary}. For digital quantum simulators it is hard to implement continuous time-dependence due to their discretized gate-based ansatz. Hence, Eq.~\eqref{eq: Kick_unitary} is implemented on digital quantum hardware to effectively perform the continuous time-dependence given in Eq.~\eqref{eq:time_ordered_integral}.

The resources required to implement a simulation via the QHiFFS algorithm depend on the forms of $\hat V(t)$ and $\hat H_0$. Here we describe a physically motivated setting in which the final simulation is particularly simple. A more detailed account of the resources required for other cases is included in~\ref{appendix: numerics}.

In what follows, we will focus on local driving (i.e., assume $\hat V(t)$ is one-local). This is a natural limit to consider since many driving phenomena, for example the optical driving of lattice systems, can be modeled in terms of local driving. In this case, $\hat K(t)$ is one-local to second order. Thus the initial and final kicks in Eq.~\eqref{eq: Kick_unitary} can be implemented using only single qubit rotations. If we further assume that $\hat V^{(j)}$ and $\hat V^{(-j)}$ commute for all $j$, which is the case here, then the first order correction to $\hat H_{\rm eff}$ vanishes and we have that $\tilde H_{\rm eff} = \hat H_{0} + O(\frac{1}{\omega^2})$.

This yields one of the main benefits of the QHiFFS algorithm, as it allows to reuse of previously developed time-independent techniques and lifts them to a Floquet Hamiltonian simulation~\cite{verstraete2009quantum, berry2015simulating,cirstoiu2020variational, gibbs2021long, mansuroglu2023variational, gibbs2024dynamical}. For example, if $\hat H_0$ is diagonal, or it can be analytically diagonalized, then one can simulate evolution under $\hat H_{\rm eff}$, and correspondingly evolution under $\hat H(t)$, using a fixed depth circuit. 
The transversely driven Ising model (independent of topology and dimension) falls into this category. 
More generally, one could use known methods for simulating the time-independent $1$D $XY$-model~\cite{verstraete2009quantum}, Bethe diagonalizable models~\cite{sopena2022algebraic} and translationally symmetric models~\cite{mansuroglu2023variational}, to simulate the effect of local driving on such systems.  

We note that in classical simulations of Floquet systems, high frequency expansions are known and commonly used~\cite{magnus1954exponential, casas2001floquet, goldman2014periodically, eckardt2015high, oka2019floquet} both for simulation and for Floquet engineering. In the latter case, Eq.~\eqref{eq:time_ordered_integral} and Eq.~\eqref{eq: Kick_unitary} are viewed from the opposite perspective to the one we take here. Namely, the time-dependent driving term, $\hat V(t)$, can be designed to implement otherwise unfeasible complex many-body time-independent Hamiltonians, $\hat H_{\rm eff}$. This perspective is also taken in the quantum computing community for gate calibration and optimization~\cite{sameti2019floquet, petiziol2021quantum, qiao2021floquet}. However, on digital quantum computers the harder task is typically to simulate the time-ordered integral. Hence, this is our focus here. Namely, we propose implementing Eq.~\eqref{eq: Kick_unitary} on quantum hardware to simulate Floquet dynamics. \ \\  

\subsection{Case study: Two-dimensional biaxial next-nearest-neighbor Ising (BNNNI) model in a transverse field}
\label{sec: results}
\begin{figure}[t!]
    \centering
    \includegraphics[width=\linewidth]{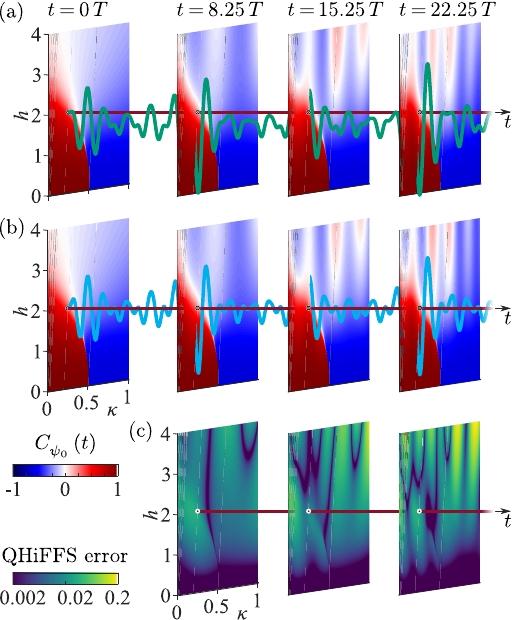}
    \caption{\textbf{Numerical simulations of non-equilibrium dynamics of the transverse field BNNNI model.}
    Two-dimensional slices of the next-nearest-neighbor correlation function, $C_{\psi_0}(t)$, as a function of $h$ and $\kappa$ for different simulation times $t$ computed exactly (a) and computed via $1^{\rm st}$ order QHiFFS (b). Here $J=1$ and $\omega=30$. The red line gives $C_{\psi_0}(t=0) \neq 0$ and the green/blue curves plot the exact/QHiFFS correlation function $C_{\psi_0}(t)$ as a function of time for the parameters $h=2$ and $\kappa=0.25$ also used in our hardware implementation (Fig.~\ref{fig: results}). 
    In (c) we plot in logscale the error in the correlation function computed via QHiFFS - that is, the difference between the slices shown in (a) and (b). Videos corresponding to this figure can be found in the \href{https://arxiv.org/src/2303.02209}{arXiv Tex Source folder}. }
\label{fig:ANNNI_numerics}
\end{figure}

\begin{figure*}[t!]
    \centering
    \includegraphics[width=\linewidth]{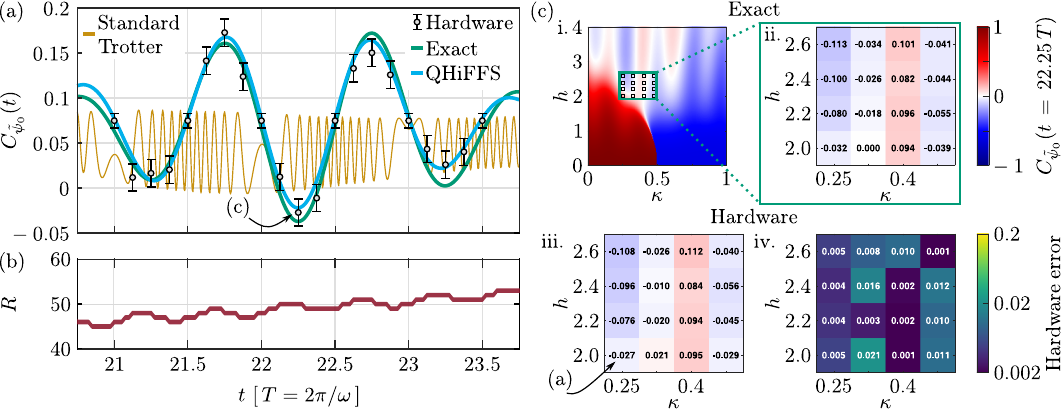}
    \caption{\textbf{Hardware results:} (a) Here we plot $C_{\tilde{\psi_0}}(t)$, Eq.~\eqref{eq:NNN},  the next-nearest-neighbor correlation function, for our short depth approximation of the ground (initial) state $\ket{\tilde{\psi_0}}$ averaged over all qubit sites $k,l$. The green line gives the exact correlation function value, the blue line is the correlation function as computed using the first-order QHiFFS ansatz, which is implementable on quantum hardware by a shallow circuit (compare Fig.~\ref{fig: quantinuum_circuit}) and the orange line a second order standard Trotterization using the same depth circuit as QHiFFS. The black data points are hardware results obtained from a 20-qubit Quantinuum H1-1 quantum computer. Their $2\sigma$-error bars  quantify  uncertainty of the results due to finite shot number. (b) The red line compares the gate overhead required by Trotterization for the same (noise-free) algorithmic fidelity. That is, we plot $ R(t) = N_{\rm Trotter}(t)/N_{\rm QHiFFS}(t)$ where $N_{\rm Trotter}(t)$ ($N_{\rm QHiFFS}(t)$) is the number of 2-qubit gates to simulate to time $t$ with a second order Trotterization (the first order QHiFFS). To make $R(t)$ a fair comparison metric, the time step for the Trotterization is set to ensure the algorithmic error is the same for both methods. (c) Here we plot the $C_{\psi}(t)$ dependence on the parameters $h$ and $\kappa$ for a zoomed in region of the total parameter space (i.) at time $t = 22.25 \,T$. (ii.) and (iii.) plot the exact correlation function values and those computed on quantum hardware respectively, with the error between them shown~in~(iv.). The raw data as well as the corresponding quantum circuits of the shown quantum hardware results in (a) and (c) are provided in the \href{https://arxiv.org/src/2303.02209}{arXiv Tex Source folder}.} 
    \label{fig: results}
\end{figure*}

The transverse field BNNNI model is an example of an Ising model with next-nearest neighbor axial interactions~\cite{elliott1961phenomenological,bak1980Ising,selke1988annni}. 
Interest in such models is largely motivated by the following key facts: Firstly, the zero-temperature critical behavior of the quantum spin Ising system in $D$-dimensions is connected to the classical critical behavior of the corresponding  $(D+1)$-dimensional classical system.  Secondly, the results of such systems can provide insight into the general role of quantum fluctuations in quantum magnetism and also have direct relevance to experiments on numerous frustrated quantum magnets~\cite{nekrashevich2022reaching}. 
While most work has focused on studying non-driven models, the non-equilibrium quantum dynamics of one-dimensional models has also been explored in a quench protocol~\cite{kennes2018controlling}. 

Here we consider the two-dimensional Floquet driven transverse field BNNNI model~\cite{hornreich1979lifschitz} defined on a square lattice with the Hamiltonian 
\begin{align}
\label{eq:BNNNI_Hamiltonian}
\hat{H}(t) = J &\left( - \sum_{\langle i,j \rangle} \hat{Z}_i  \hat{Z}_j + \kappa \sum_{\langle\langle i,j \rangle\rangle}  \hat{Z}_i  \hat{Z}_j\right) \\
&- h \cos(\omega t) \sum_i  \hat{X}_i \nonumber \, .
\end{align}

Here $\hat{Z},\hat{X}$ are Pauli operators, $\sum_{\langle i,j \rangle}$ denotes a sum over the nearest neighbors, and $\sum_{\langle \langle i,j  \rangle\rangle}$ is a sum over axial next-nearest neighbors as shown in  Fig.~\ref{fig:ansatz_idea}(b). For simplicity, below we consider only the case of $J,\kappa>0$, for which the model is a paradigmatic example of frustrated magnetism.

Classical simulations of two-dimensional frustrated magnets are very challenging. Even their equilibrium properties are subject to long-standing controversy~\cite{savary2016quantum}. In general, classically simulating their non-equilibrium properties is even more challenging, which has stimulated widespread interest in their quantum simulation~\cite{eisert2015quantum}. These challenges also limit our understanding of physics of the transverse field BNNNI model.
Its static phase diagram is well-established for special cases.  The zero-field ($h = 0$) system exhibits a phase transition from a ferromagnetic phase to an antiphase at $\kappa_c=\frac{1}{2}$~\cite{hornreich1979lifschitz}. In the ferromagnetic phase, all spins point in the same direction, while in the antiphase one finds periodic sequences of $2$ spins pointing in the same direction, followed by $2$ spins pointing in the opposite direction~\cite{hornreich1979lifschitz}.  The presence of a transverse magnetic field, $h\hat{X}$, introduces additional quantum fluctuations. For $\kappa=0$, we recover the extensively studied transverse field quantum Ising model~\cite{pfeuty1970one, pfeuty1971ising} with ferromagnetic order for $h<h_c\approx3.04$~\cite{blote2002cluster}. At $h_c$ the model undergoes a  phase transition to a paramagnetic phase with all spins aligned in the $\hat{X}$ direction in the limit of $h\gg1$. 

To probe dynamical properties and nonequilibrium phases of this model we study the next-nearest-neighbor correlation function averaged over all qubits on the lattice. That is, for an $n_x\times n_y$ lattice, we compute
\begin{equation}
    C_{\psi}(t) := \frac{1}{n_x n_y} \sum_{l_x=0}^{n_x-1} \sum_{l_y=0}^{n_y-1} \Braket{\psi | \hat U(t)^\dagger \hat{Z}_{l_x,l_y}\hat{Z}_{l_x,l_y+2} \hat U(t) | \psi} 
    \label{eq:NNN}
 \end{equation}
where $\ket{\psi}$ denotes an initial state of interest. To fit the lattice onto Quantinuum’s H1-1 quantum computer, in our hardware implementation, we consider a $4\times5$ lattice with periodic boundary conditions. The correlator $ C_{\psi}(t)>0$ indicates that directions of the next-nearest axial neighbor spins are aligned as in the  ferromagnetic phase,  while  $ C_{\psi}(t)<0$ indicates that they are anti-aligned (as in the antiphase). This quantity can serve to probe the field-induced phase change. In our case, we generally take $\ket{\psi}$ to be the ground state, denoted by $\ket{\psi_0}$ of  $\hat H(t=0)$, setting $t_0=0$ in the following. The choice sets our simulations realistically connected with experimental situations. We note that since $C_{\psi}(t)$ is computed in Eq.~\eqref{eq:NNN} as a sum of expectation values of observables diagonal in the computational basis, we use a computational basis measurement to estimate all of them from each shot  maximizing shot-efficiency of $C_{\psi}(t)$ estimation.   

\subsubsection{Numerical Simulations}\label{subsec: numeric results}

We performed numerical simulations to study how well the QHiFFS algorithm captures short range correlation functions of the periodically driven BNNNI model. Figure~\ref{fig:ANNNI_numerics}(a) shows the next-nearest-neighbor correlation function, $C_{\psi_0}(t)$, as a function of $h$ and $\kappa$ for different simulation times. We chose $\omega=30$, which ensures the applicability of the kick approximation for all considered parameter values of the model. Regions of ferromagnetic and antiferromagnetic  correlation are indicated in red and blue respectively. 

We see that periodic driving further enriches the physics of the BNNNI model. 
In particular, when the BNNNI system is driven out of equilibrium by a high-frequency high-strength transverse field, a noticeable unstable ferromagnetic tendency can be induced out of an otherwise antiphase near the $\kappa=\frac{1}{2}$ quantum critical fan. While this instability persisted for all times we studied (i.e., until the limits of our first-order QHiFFS simulation), we expect that this is a transient feature of a pre-thermalization phase. Physically, it can be understood as follows: In the non-driving case, the system stays in one of the three major equilibrium states (ferromagnetic, antiphase, or paramagnetic in the $h$-$\kappa$ nonthermal parameter space. The low-$h$ Floquet BNNNI system maintains the dominant feature of its original equilibrium magnetic state (ferromagnetic or antiphase). When the magnetic field amplitude becomes sufficiently strong such that the equilibrium state is in the paramagnetic phase, the driven Floquet system effectively experiences a sequence of large and low field. As such, the equilibrium paramagnetic regime is now replaced one carrying the feature of low-field ordered phases. A similar behavior has also recently been discovered in a heavy-fermion Kondo lattice in a classical simulation~\cite{fauseweh2020laser}. Although a determination of Floquet dynamical state requires a full evaluation of spatially dependent spin-spin correlator, our result suggests that driven quantum systems are an exciting research area for generating and manipulating quantum phases.

Importantly, these complex changes are well-captured by the QHiFFS algorithm. This is shown in Fig.~\ref{fig:ANNNI_numerics}(b) where we plot the correlation function, $C_{\psi_0}(t)$, as computed using the QHiFFS algorithm. Indeed, as shown in Fig.~\ref{fig:ANNNI_numerics}(c), the errors in the correlation function (that is, the difference between the exact correlation function and the one computed via QHiFFS) are small. Specifically, the correlation function error on average (over the different $h$ and $\kappa$ values) is 0.0121 at $t = 8.25\,T$, 0.0175 at $t = 15.25\,T$ and 0.0216 at $t = 22.25\,T$. In \ref{appendix: numerics}, we provide equivalent numerical results on the $4 \times 6$ and $5 \times 5$ transversely driven BNNNI model; indicating that these larger versions could have also worked on quantum hardware. \ \\

\subsubsection{Hardware Implementation}\label{subsec: hardware results} 
To demonstrate the suitability of the QHiFFS algorithm for real quantum hardware we used all 20 qubits of Quantinuum's H1-1 trapped ion platform to simulate the $4 \times 5$ BNNNI model with periodic boundary conditions (see Fig.~\ref{fig:ansatz_idea} and for the executed hardware circuit see Fig.~\ref{fig: quantinuum_circuit}). This was enabled by the all-to-all connectivity of the platform -- on superconducting circuit devices (as compared to the trapped ion device that we used here), periodic boundary conditions on two-dimensional lattice models are typically hard to implement due to the restriction to nearest-neighbor gates. 

Due to a finite sampling budget of approximately 30,000 shots, we needed to restrict ourselves to specific model parameters. We decided to implement two studies - the first a simulation of the correlation function as a function of time, the second a slice of the correlation function at a fixed time plotted versus the model parameters. 
Without loss of generality, we took $J=1$. For our plot of the correlation function as a function of time, we then chose $h=2$, $\kappa = 0.25$, and $\omega=30$ for which our numerical results indicate strong effects of periodic driving. For our study of model parameter dependence, we took $t = 22.25\, T$ and focused on the region $0.25 \leq \kappa \leq 0.475 $ and  $2.0 \leq h \leq 2.6 $.
The chosen parameters ensured that all model parameters are of a similar order of magnitude resulting in strong frustration and quantum fluctuations. 

In Fig.~\ref{fig: results}(a) we plot as a function of time: the correlation function obtained from our quantum hardware implementation (black data points), the exact value of the correlation function (green line) and the value predicted from classical simulations of the QHiFFS algorithm (blue line). Furthermore, for the hardware results we provide error bars quantifying their uncertainty due to a finite shot number as two standard deviations of the sample mean. Fig.~\ref{fig: results}(b) shows the ratio of the number of 2-qubit gates that would have been required for a Trotter simulation as compared to a QHiFFS simulation for the same average simulation fidelity.
This varied from 45-fold (at short times) to 53-fold (at long times) and thus the Trotter simulation was not feasible to be implemented on quantum hardware.

In Fig.~\ref{fig: results}(c) we plot the zoomed in  $C_{\psi}(t)$ dependence on the parameters at time $t = 22.25 \,T$ computed exactly and as computed on quantum hardware. The hardware error is less than 0.021 for all data points. Our implementation successfully captures the noticeable unstable ferromagnetic tendency induced out of the antiphase at the $\kappa=\frac{1}{2}$ quantum critical fan. This suggests that QHiFFS could open up new avenues to study novel driven quantum phases and criticality in strongly correlated electron systems.
This could become particularly relevant experimentally when material systems with interesting high energy scales are subject to an electromagnetic field.

The expected values of the correlation function are largely within our shot-noise based error bars, implying that our simulation (despite the considerable circuit depth and qubit number) is predominantly shot-noise limited. Indeed, the simulation is also limited by the size of the available hardware - we chose to study a 20 qubit simulation not because this was the largest QHiFFS could handle, but rather because the Quantinuum H1-1 system was the largest all-to-all device to which we had access. Due to this shot noise limit, we expect the further exemplary models we studied in Sec.~\ref{subsec: 2-local models} and \ref{appendix: numerics} would work similarly well. \ \\

\subsubsection{Error Analysis}
\label{subsec: error estimate}
\begin{figure*}[t!]
    \centering
    \includegraphics[width=\linewidth]{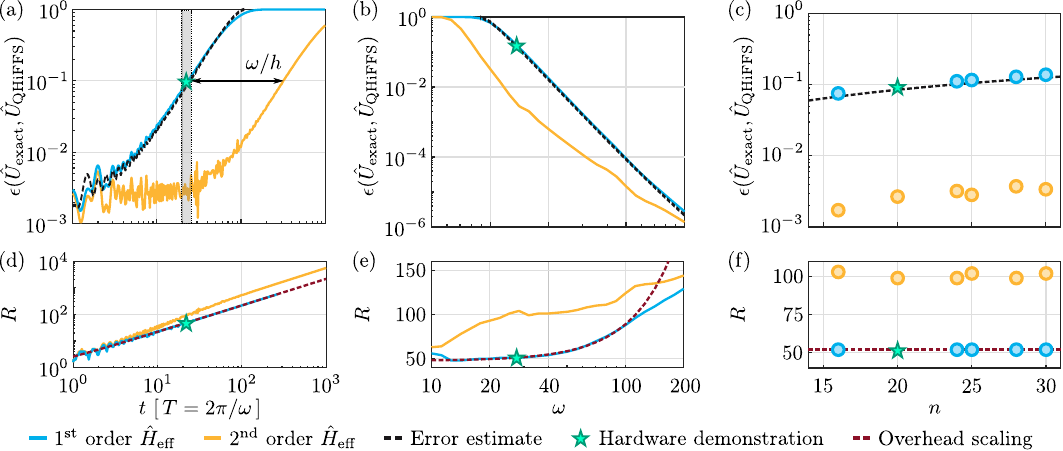}
    \caption{\textbf{Error analysis.} 
    Here we plot the average fidelity as a function of (a) simulation time $t$, (b) frequency $\omega$ and (c) system size $n$. Unless varied, the parameters equal those of our hardware implementation in Fig.~\ref{fig: results}; namely $J\,=\,1$, $\kappa\,=\,0.25$, $h\,=\,2$, $\omega\,=\,30$ and $t\,=\,22.75\,T$. Overall, we find an excellent agreement between our analytical error estimate (black) and the exact numerical calculations (cyan). (Note, we plot here the full 1st-order expression for the error given in Eq.~\eqref{eq:error_ANNNI2D} in \ref{app:derivation}). The yellow line indicates the exact error when $\hat{H}_{\rm eff}$ is expanded to second order in $\omega$ and the kick operator is expanded to first order in $\omega$. (d), (e) and (f) show the numerically calculated overhead of standard Trotterization $R$ in terms of simulation time $t$, frequency $\omega$ and system size $n$. For first order QHiFFS (cyan) we observe linear, quadratic (since $t$ is chosen in units of $T=\frac{2\pi}{\omega}$) and constant scalings with $t$, $\omega$ and $n$ respectively (shown in red) as predicted by  Eq.~\eqref{eq:error_comparision}.}
    \label{fig:error_scaling}
\end{figure*}

To place the algorithm on solid conceptual foundations, as well as better understand the results of our numerical simulations and hardware implementations, we have conducted an analysis of QHiFFS algorithmic errors. 
To quantitatively judge the quality of a given Floquet simulation, we consider the average simulation infidelity over the uniform distribution of input states
\begin{align}
\epsilon(\hat U_1, \hat U_2) := 1 - \mathbb E \left. \left[ \left| \Braket{\psi |\hat U_1^\dagger \hat U_2|\psi} \right|^2\right]\right|_{\ket \psi \sim \mathrm{Haar}_n} \, .
 \label{eq:error_def}
\end{align}
Here we provide a summary of our error analysis for the transversely driven BNNNI model. In \ref{appendix: kick_error} and \ref{appendix: trotter_error}, we frame our error analysis more generally.

We start by analyzing the scaling of errors for the standard Trotterization approach. As shown in detail in~\ref{appendix: numerics} and \ref{appendix: trotter_error}, there are three sources of error in this case, one coming from the discretization of the time-ordered integral, one coming from the non-commutativity of the Hamiltonian at different times and another (the standard time-independent Trotter error) coming from the non-commutativity of terms in the Hamiltonian at a fixed time. The dominant contribution in the high-frequency regime comes from discretization, while the errors from non-commutativity are comparably small. We find that the average infidelity for the transverse BNNNI model scales as
\begin{align}
    &\epsilon(\hat U_{\textrm{exact}} (t), \hat U_{\textrm{Trotter}} (t)) = \mathcal{O} \left( \frac{n h^2 t^4 \omega^2 }{8 m^2} \right).
\label{eq:Trotter_error}
\end{align}
where $m$ is the number of Trotter steps and $n$ is the system size. Thus, in contrast to standard time-independent Trotterization, the error here is independent of $\hat H_0$ and scales as $t^4$ rather than $t^2$.

In comparison, we find that the long time error for the transversely driven BNNNI model when simulated by QHiFFS, can be approximated as
\begin{align}
&\epsilon(\hat U_{\textrm{exact}}(t), \hat U_{\textrm{QHiFFS}}(t)) \approx  \frac{281}{32}\frac{n\cdot h^4 \cdot (1+\kappa^2) \cdot J^2 \cdot t^2}{\omega^4} \, .
\label{eq:error_ANNNI2D}
\end{align}
The dominant contribution here is from the high-frequency truncation of the effective Hamiltonian. At short times, the truncation of the Kick approximation also contributes a significant oscillatory error to the averaged infidelity. However, the relative significance of this effect is washed out at longer times due to the contribution from the effective Hamiltonian which grows quadratically in time (compare~\ref{app:derivation}). We also provide a rigorous upper bound for $\epsilon(\hat U_{\textrm{exact}}(t), \hat U_{\textrm{QHiFFS}}(t))$ in~\ref{appendix: kick_error} and show that it admits the same scaling in $t$ and $\omega$ as Eq.~\eqref{eq:error_ANNNI2D}, namely is in
$\mathcal{O}\left(\frac{t^2}{\omega^{4}}\right)$.

To support our analysis, we numerically computed the exact errors for the transversely driven BNNNI model. Specifically, in Fig.~\ref{fig:error_scaling} we plot these errors, and our corresponding error estimates for the parameter regime we simulated on quantum hardware. Namely, a $4 \times 5$ two-dimensional transversely-driven BNNNI model with $J=1$, $\kappa=0.25$, $h=2$, $\omega = 30$ (in (a)) and $t = 22.75 T$ (in (b)). Overall, we find an excellent, nearly indistinguishable, agreement with our analysis.

Our error analysis further allows us to estimate how the circuit depth of a standard Trotterization needs to scale compared to QHiFFS to achieve the same fidelity. Let $R$ be the ratio of the number of two qubit gates used by Trotterization compared to the number of two qubit gates used by QHiFFS. For the driven BNNNI model presented here $R = m$, the number of Trotter steps used by the Trotterization since the $1^{\rm st.}$ order QHiFFS requires the same number of two qubit gates as one ($1^{\rm st.}$~or $2^{\rm nd.}$ order) Trotter step.  
Specifically, comparing our upper bound Eq.~\eqref{eq:Trotter_error} and estimate Eq.~\eqref{eq:error_ANNNI2D} we would expect $R$ to scale as
\begin{align}
    R \sim  \frac{t \omega^3}{J h \sqrt{1+\kappa^2}} \, ,
    \label{eq:error_comparision}
\end{align}
assuming our bound and estimate are tight.
It turns out that this predicted scaling does indeed well-reproduce the exact scaling for $R$ computed numerically. Namely, the Trotter overhead grows linearly in time, as shown in Fig.~\ref{fig: results}(b) and Fig.~\ref{fig:error_scaling}(d), and cubically in $\omega$,  as shown in Fig.~\ref{fig:error_scaling}(e).
We stress that this observation is independent of system size, $n$, as shown in Fig.~\ref{fig:error_scaling}(f).

The advantage of the QHiFFS ansatz can be further boosted by truncation at higher orders in $k$, (i.e.  $(1/\omega)^k$). If we truncate at order $k$, the ratio of circuit depths scales instead as 
\begin{align}
R \sim \frac{t \omega^2}{J \sqrt{1+\kappa^2}} \left(\frac{\omega}{h} \right)^k \, .
 \label{eq:error_comparision_higher_w}
\end{align}
Such higher order expansions open up the simulation of longer times and larger system sizes without increasing the algorithmic approximation infidelity, giving a significant advantage to the QHiFFS approach. 

\subsection{Additional Models}
\label{subsec: 2-local models}

In \ref{appendix: numerics} we provide additional numerics to support the broad applicability of QHiFFS. In order to discuss an entangling driving term that leads to faster growth of locality in $\hat H_{\rm eff}$ as well as an effective time-independent Hamiltonian that needs to be trotterized, we analyze two more models each exhibiting at least one of these features. To be precise, we study a model that is derived from the BNNNI Hamiltonian, Eq.~\eqref{eq:BNNNI_Hamiltonian}, by substituting the non-entangling driving term with the following 2-local term 
\begin{align}
   \hat V(t) = - J_X \cos(\omega t) \sum_{\langle i,j \rangle}  \hat{X}_i\hat{X}_j.
\end{align}
Following extensive numerical studies in~\ref{appendix: numerics_BNNNI_XX}, we find no noteworthy difference in the QHiFFS performance of 2-local compared to 1-local driving. In Fig.~\ref{fig: observables_BNNNI_ZZ_cos_wt_XX} we provide examples of the time-evolved correlators similar to Fig.~\ref{fig: results} (a) and (b), and in Fig.~\ref{fig:error_scaling_ANNNI_cos_wt_XX}  and Fig.~\ref{fig:error_scaling_ZZ_cos_wt_XX} we show the algorithmic error scaling (equivalently to Fig.~\ref{fig:error_scaling}).

Further, we consider a driven XY model of the following form
\begin{align}
    \hat{H}(t) = &-J_X  \sum_{\langle i,j \rangle} \hat{X}_i  \hat{X}_j -J_Y  \sum_{\langle i,j \rangle} \hat{Y}_i  \hat{Y}_j \nonumber \\
    &- J_Z \cos(\omega t) \sum_{\langle i,j \rangle}  \hat{Z}_i\hat{Z}_j.
\end{align}
In this case the truncated effective Hamiltonian $\tilde H_{\rm eff}$ also needs to be trotterized. We show, that also in these cases QHiFFS yields an advantage over the Trotterization of the time-ordered exponential (see \ref{appendix: numerics} and \ref{appendix: kick_error}). The reason for this improvement lies in the error sources of the Trotterization of time-ordered exponentials. Besides the error from non-zero commutation of terms in $\tilde H_{\rm eff}$ form which both QHiFFS and naive Trotterization suffer, the latter also acquires errors additionally from non-zero commutation of $\hat H(t)$ at different times, errors from discretizing the integrals, as well as non-suppressed commutator errors between $\hat H_0$ and $\hat V(t)$; compare Eq.~\eqref{eq:QHiFFS_bound} (QHiFFS) and Eq.~\eqref{eq:Trotter_error_full} (Trotter) as well as Eq.~\eqref{eq:appendix_general_QHiFFS_error} and Eq.~\eqref{eq:appendix_general_Trotter_error}. Equivalently to the other two numerical examples, we give examples of (QHiFFS) time evolved observable in Fig.~\ref{fig: observables_XX_YY_cos_wt_ZZ_small_Jxy} and Fig.~\ref{fig: observables_XX_YY_cos_wt_ZZ_Jxyz_1}, including an illustrative explanation why standard Trotterization breaks for single frequency driving within the first half period. Analogously, the algorithm scaling behavior is shown in Fig.~\ref{fig:error_scaling_XX_YY_cos_wt_ZZ} matching our predictions in Eq.~\eqref{eq:appendix_e13} to Eq.~\eqref{eq: Std_Trotter_norm_XX_YY_cos_wt_ZZ}. Qualitatively, we find that the QHiFFS error is then typically dominated by the BCH-error of the effective Hamiltonian; while QHiFFS maintains an error scaling advantage given by the avoided time-independent BCH-error and shorter Trotter sequence of $\hat H_\text{eff}$ compared to $\hat H(t)$ (e.g. see Eq.~\eqref{eq: R_XX_YY_cos_wt_ZZ} and compare Eq.~\eqref{eq:appendix_e13} with Eq.~\eqref{eq:appendix_e14}). We discuss this in more detail in the Appendix.

\section{Discussion}
\label{sec:discussion}

Here we have proposed the QHiFFS algorithm for simulating periodically driven systems on digital quantum computers. Central to the algorithm is the use of a kick operator to transform the problem into a frame of reference where the dynamics of the system is governed by a time-independent effective Hamiltonian. Thus, this method avoids the costly discretization of the time-ordered integral required by standard Trotterization and allows one to re-use previous time-independent Hamiltonian simulation methods~\cite{verstraete2009quantum, berry2015simulating,cirstoiu2020variational, gibbs2021long, mansuroglu2023variational, gibbs2024dynamical}. In parallel, QHiFFS does not require any form of variational optimization of the sort required by many other near-term simulation methods~\cite{cirstoiu2020variational,gibbs2021long,gibbs2024dynamical,mansuroglu2023variational}.

Our error scaling analysis demonstrates that QHiFFS outperforms standard Trotterization for high driving frequencies, $\omega$. Specifically, at long simulation times, $t$, we see a linear advantage in $t$ and at least cubic improvements in $\omega$ compared to standard Trotterization, for diagonalizable effective Hamiltonians. In parallel, the algorithmic errors of QHiFFS scale linearly in system size, $n$, for local models, and at worst quadratically. These favorable scalings again point towards QHiFFS's suitability for larger-scale implementations.

To benchmark QHiFFS we performed a 20-qubit simulation of the periodically driven two-dimensional BNNNI model on Quantinuum's 20-qubit H1-1 trapped ion platform. As shown in Fig.~\ref{fig: results}, the correlation function values computed on the quantum computer capture the true predicted values up to the precision allowed by shot noise ($2\sigma = 1.49 \cdot 10^{-2}$). In contrast, standard Trotterization (specifically a 2$^{\rm nd}$ order Trotterization with the same algorithmic error as a 1$^{\rm st}$ order QHiFFS) would have required approximately 50 times as many two-qubit gates, making the method completely unfeasible on current hardware due to gate errors and decoherence. 

Future interesting applications of QHiFFS on quantum hardware could include the simulation of strongly correlated systems in the presence of linearly or circularly polarized electromagnetic field for the study of the emergence of unconventional superconductivity~\cite{kumar2021inducing}, quantum spin liquids~\cite{claassen2017dynamical}, and Kondo coherence collapse~\cite{zhu2021ultrafast}. Similarly, one could study even more complicated, approximately fast-forwardable Ising models, either by adding longer range interactions or by studying three spatial dimensions. We believe the latter would be viable with current gate fidelities, e.g. on Quantinuum's trapped ion device, but (for a non-trivial simulation) would require more qubits.

In practice, QHiFFS shows to be suitable to simulate strongly correlated insulating systems, in which the charge degree of freedom is fully gapped out to a much larger energy scale while the spin becomes the only low-energy degree of freedom. In these systems, the typical exchange interaction is at an energy scale of 10 meV. As such, the frequency of an electromagnetic field can be easily chosen to be much bigger than the exchange interaction strength.

In special cases, namely those where the time-independent Hamiltonian $\hat H_0$ and the driving term $\hat V(t)$ are diagonal or efficiently diagonalizable, a fixed depth circuit can be used to simulate arbitrary times with an algorithmic error that grows only quadratically in time; compared to the power of 4 for standard Trotterization. Thus, in this limit, the QHiFFS algorithm provides a new method for approximate \textit{fast-forwarding} quantum simulations~\cite{cirstoiu2020variational, gibbs2021long, gibbs2024dynamical} that is highly suitable to near-term hardware. 

We stress that while the simulation in such cases is particularly simple, and thus also simpler to simulate classically, QHiFFS could still be used to perform classically intractible simulations. This is because even the fixed depth circuits enabled by the kick approximation could be challenging to simulate classically. In general, apart from special cases like Clifford circuits or one-dimensional low entangled states~\cite{verstraete2008matrix}, the classical cost of simulating quantum circuits grows exponentially with circuit depth. Therefore, one may expect that sufficiently deep and entangling fixed-depth QHiFFS circuits will be challenging to simulate classically in higher spatial dimensions. Given that these gate sequences are good approximations of the dynamics of quantum many-body systems that are forced out of equilibrium by a high-frequency drive, such strong entanglement generation is a natural expectation. Furthermore, even in the case where QHiFFS is classically simulable for some input states, it could still be used to implement classically intractable simulations for non-classically simulable initial states. Such states could plausibly be prepared via analog simulation strategies. This could provide one of the earliest avenues for physically interesting non-classically simulable implementations of QHiFFS on near-term hardware.

Longer term, we expect QHiFFS to find use in the fault-tolerant era as well as the near-term era. In the former case, one could use higher order expansions of the effective Hamiltonian and kick operators to achieve a higher accuracy, and implement the more complex resulting effective Hamiltonian using well-established fault-tolerant simulation methods~\cite{berry2015simulating}.

\begin{figure*}[t!]
    \centering
    \includegraphics[width=\linewidth]{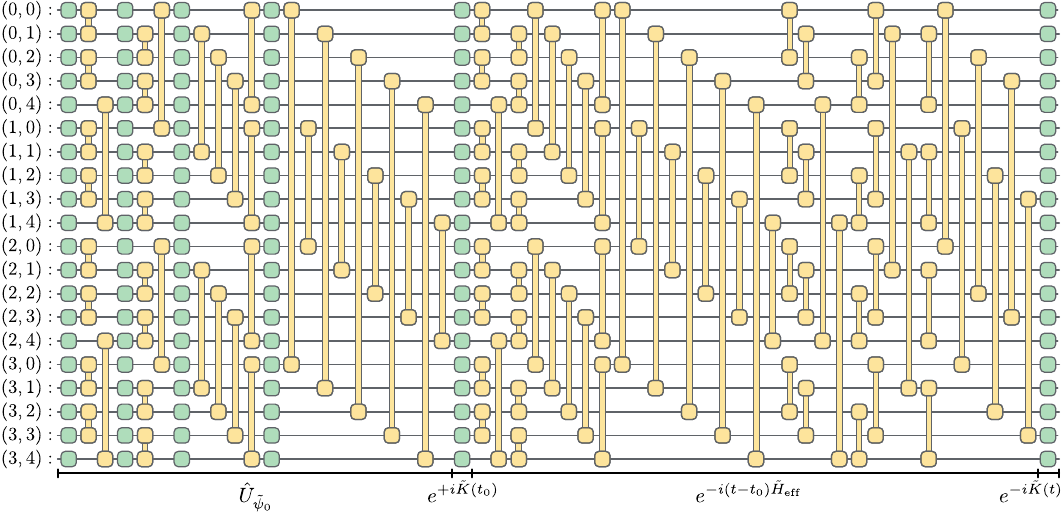}
    \caption{\textbf{Quantum hardware circuit:} Native universal one-qubit gates are colored in green and native $RZZ$ two-qubit gates are depicted in yellow. 
    The first 4 two-qubit layers (40 gates) are used for approximate preparation of the BNNNI model ground-state $\ket{\psi_0}$ (average overlap $\approx 96.4$~\%). The remaining 7 two-qubit layers (70 gates) are used to implement $\exp (-i (t- t_0) \hat H_{\rm eff}) \approx \exp (-i(t-t_0) \hat H_{0})$. Both $\hat K(t)$ and $\hat H_0$ can be implemented exactly up to $\mathcal O \left (\frac{1}{\omega^2}\right)$ (compare Eqs.~\eqref{eq: Heff_series_1st_order} and \eqref{eq: Kick_series_1st_order}), which allows for constant depth approximate fast-forwarding. } 
    \label{fig: quantinuum_circuit}
\end{figure*}

\section{Methods}
\subsection{Algorithmic Methods}
Here we provide an overview (for details see \ref{app:derivation}) on how to obtain an effective time-independent Hamiltonian $H_{\rm eff}$ for Floquet simulation. Using both Schrödinger equations (for $\hat H(t)$ and $\hat H_{\rm eff}$) and a time dependent transformation $e^{-i \hat K(t)}$ one gets a defining relation for $\hat H_{\rm eff}$,
\begin{align}
    \hat H_{\rm eff} = e^{i\hat K(t)} \hat H(t) e^{-i \hat K(t)} + i \left( \frac{\partial e^{i \hat K(t)}}{\partial t} \right) e^{-i \hat K(t)} \,. \label{eq:HiF_def}
\end{align}
Using the Hadamard Lemma and $\frac{\partial}{\partial x} e^{\hat A(x)} = \int_0^1 e^{y \hat A(x)} \left( \frac{\partial}{\partial x} \hat A(x) \right) e^{(1-y) \hat A(x)} dy$ this can be rewritten in nested commutator form:
\begin{align}
    \hat H_{\rm eff} &= \sum_{q = 0}^\infty \frac{i^q}{q!} \comm{\hat K(t)_{(q)}}{\hat H(t)} \nonumber \\
    &-\sum_{q=0}^\infty \frac{i^{q}}{(q+1)!} \comm{\hat K(t)_{(q)}}{\frac{\partial \hat K(t)}{\partial t}}  
    \label{eq:HiF_def2}
\end{align}
with $\comm{\hat K(t)_{(q)}}{\hat H(t)} = \comm{\hat K(t)}{\comm{\hat K(t)_{(q-1)}}{\hat H(t)}}$. So far, no structure of the problem or approximation is used. In the case of periodic time-dependence, Floquet's theorem implies one may use a high-frequency expansion for the kick operator and effective Hamiltonian,
\begin{align}\label{eq:perturbexpan}
    \hat K(t) = \sum_{k=1}^{\infty} \frac{1}{\omega^k} \hat K^{(k)}(t) \quad \mathrm{and} \quad
    \hat H_{\rm eff} =\sum_{k=0}^{\infty} \frac{1}{\omega^k} \hat H_{\rm eff}^{(k)} \, ,
\end{align}
and truncate at (small) finite order~\cite{goldman2014periodically, goldman2015erratum, eckardt2015high}.
We note that for a different time-dependence structure another ansatz for $\hat K(t)$ and $\hat H_{\rm eff}$ might lead to preferable gate sequences. Given a specific ansatz, such as the high-frequency expansion in Eq.~\eqref{eq:perturbexpan}, the truncated series can be determined explicitly by the following algorithm. Going through these steps for $k_{max}=1$ gives Eqs.~\eqref{eq: Heff_series_1st_order} and~\eqref{eq: Kick_series_1st_order}.

\RestyleAlgo{ruled}
\RemoveAlgoNumber
\begin{algorithm}[hbt!]
\caption{for deriving $\tilde K(t)$ and $\tilde H_{\rm eff}$}
$k \gets 0$\;
\While{$k < k_{max}$}{
    \quad Consider Eq.~\eqref{eq:HiF_def2} up to order $k$\;
    \quad Plugin $\forall j \in \{1,...,k \}:\, \hat K^{(j)}(t)$ and $\hat H_{\rm eff}^{(j-1)}$ \;
    \quad $\hat K^{(k+1)}(t) \gets $ time-dependent terms\;
    \quad $\hat H_{\rm eff}^{(k)} \gets$ time-independent terms\;
    \quad $k \gets k + 1$\;
}
\end{algorithm}

\subsection{Simulation Methods}
The first order QHiFFS ansatz yielded a constant depth circuit to simulate the BNNNI model on quantum hardware as both $\exp (-it \tilde H_\mathrm{eff})$ and $\exp (\pm i \tilde K(t))$ were exactly implementable. Indeed, in the case of Quantinuum's H1-1 trapped ion device, with native single qubit rotations and two-qubit gates $\exp (-i \frac{\theta}{2} \hat{Z}\hat{Z})$, the simulation was implemented exactly with a fixed circuit using the native gate set. This meant that the QHiFFS circuit, for any simulation time, had the same depth as a single standard low-order Trotter-Suzuki step. Thus, the final simulation time $t$ affected the error from the high-frequency approximation, but not the error from hardware imperfections. 

Initial state preparation for a physically motivated quantum simulation is a non-trivial task, in general. In our case, preparing the ground states exactly would require deep circuits but using classical compilation techniques we found an approximation of the ground state that still captured the same underlying physics and required a much shorter depth circuit to prepare. Our final circuits used $40$ $2$-qubit gates to prepare an approximation of the ground state for different $h$ and $\kappa$ with an average overlap of $0.964$ relative to the true ground state.

The complete quantum circuit for the full simulation consisted of $110$ $2$-qubit gates and $120$ single qubit gates (see Fig.~\ref{fig: quantinuum_circuit}) for all non-half-integer time steps. At half-integer time steps the circuit implementing the time evolution compiled to the identity and so the circuit consisted of only the 40 $2$-qubit gates required for state preparation. As an example, we provide the precise hardware quantum circuit for $h=2$, $\kappa=0.25$ and $t = 22.25\,T$ (compare Fig.~\ref{fig: results}) as well as the code to run it via Quantinuum's API in the \href{https://arxiv.org/src/2303.02209}{arXiv Tex Source folder}.

Classical numerical simulations in Fig.~\ref{fig:ANNNI_numerics}~-~\ref{fig:error_scaling} were performed using a state vector simulator. To simulate the exact time evolution, we trotterized it and decomposed the resulting evolution operator to single- and two-qubit unitaries enabling the usage of the state vector simulator. The Trotter step was determined using analytic bounds on the Trotterization error discussed in the following Sec.~\ref{sec:meth_analytics}.   \\

\subsection{Analytical Methods}
\label{sec:meth_analytics}
Upper bounds for QHiFFS and standard Trotter decomposition are derived using the estimate:
\begin{align}
    \norm{\prod_{j=1}^{n_j} e^{i \hat A_j} - \prod_{j=1}^{n_j} e^{i \hat B_j} } \leq \sum_{j=1}^{n_j} \norm{\hat A_j - \hat B_j}.
    \label{eq:bound}
\end{align}
For the QHiFFS approximation ($n_j=3$) this allows to derive a general upper bound by comparison of different orders (details see~\ref{appendix: kick_error}):
\begin{align}
    &\norm{\hat U_{\rm exact}(t_0, t) - \tilde U_{\rm QHiFFS}(t_0, t)} \nonumber \\&\leq \norm{\hat K(t) - \tilde K(t)} + |t-t_0| \norm{\hat H_{\rm eff} - \tilde H_{\rm eff}} \nonumber \\
    &+ \norm{\hat K(t_0) - \tilde K(t_0)} 
    = \mathcal{O}\left( \frac{|t-t_0|}{\omega^{k+1}} + \frac{1}{\omega^{k+1}} \right),
    \label{eq:QHiFFS_bound}
\end{align}
where $\hat \square$ indicates exact expansions and $\tilde \square$ labels the truncation at $1/\omega^k$. This means we can upper-bound the QHiFFS error by simply adding the individual errors in both kick operators and the effective Hamiltonian. 
To derive an accurate estimate of the QHiFFS error, the norm $\norm{\hat U_{\rm exact}(t) - \tilde U_{\rm QHiFFS} (t)}$ can be expanded in its trace form $\text{Tr}(\hat U_{\rm exact}(t)^\dagger \tilde U_{\rm QHiFFS} (t))$ and in orders of $\frac{1}{\omega}$ instead of bounding it. To do this, the truncated (e.g. $e^{-i \tilde K(t)}$) and the non-truncated exponential (e.g. $e^{-i \hat K(t)}$) are merged via the identity
\begin{align}
    e^{\hat A} e^{-\hat A +\chi} - \mathds{1} = \sum_{j=0}^{\infty} \frac{1}{(j+1)!} \comm{\hat A^{(j)}}{\chi} + \mathcal{O}(\chi^2),
\end{align}
which is a combination of the Hadamard lemma, the Baker-Campbell-Hausdorff lemma and a general formula for the derivative of the exponential map. $\chi$ denotes the tail of the high-frequency expansion and is small compared to $\hat A$. The leading error term is hence determined by the linear contribution in $\chi$ (see~\ref{app:derivation} for details). Explicitly doing this calculation for the BNNNI model gives Eq.~\eqref{eq:error_ANNNI2D}.

To derive the standard Trotter error bound and estimate respectively, we can use Eq.~\eqref{eq:bound} to upper bound the standard Trotterization error
\begin{align}
        &\left| \left|  \hat U_{\rm exact}(t_0, t) - \hat U_{\rm Trotter}(t_0, t) \right| \right| \nonumber \\
        &\leq \left| \left| \mathcal{T} e^{-i \int_{t_0}^{t} \hat H(s) ds } - e^{-i \int_{t_0}^{t} \hat H(s) ds } \right| \right| \nonumber \\
        &+ \left| \left| e^{-i \int_{t_0}^{t} \hat H(s) ds }  - \hat U_{\rm discrete}(t_0, t) \right| \right| \nonumber \\
        &+ \left| \left| \hat U_{\rm discrete}(t_0, t) - \hat U_{\rm Trotter}(t_0, t) \right| \right| 
        \label{eq:Trotter_error_full}
    \end{align}
with $ \hat U_{\rm discrete}(t_0, t) = \prod_{r=1}^{m} e^{i\delta t \hat H(t_0 + r\delta t)}$ and $\hat U_{\rm Trotter}(t_0, t) = \prod_{r=1}^{m} \prod_{j=1}^{n_j} e^{i\delta t \hat H_j (t_0 + r\delta t)}$. The first term describes errors due to non-commutativity of $\hat H(t)$ at different times, the second term gives the integral discretization error and the third term is standard non-time-dependent Trotter error.

For the BNNNI model the first and third term scale as $\mathcal{O}\left( \frac{t^2}{m} J(1+\kappa) h \right)$, which is why the dominant error term arises from the Riemann discretization of the time integral, which can be bounded by another application of Eq.~\eqref{eq:bound},
\begin{align}
		&\left| \left| e^{-i \int_{t_0}^{t} \hat H(s) ds } - \hat U_{\rm discrete}(t_0, t) \right | \right | \leq \nonumber \\
  &\left|\left| \int_{t_0}^{t}  \hat H(s) ds - \delta t \sum_{r=1}^m \hat H( t_0 + r \delta t)\right|\right| \, .
        \label{eq:Trotter_bound}
	\end{align} 
 Explicitly calculating this then gives the error estimate in Eq.~\eqref{eq:Trotter_error} (compare~\ref{appendix: trotter_error}).

 \section*{Code availability}
The quantum hardware circuits which were run on a Quantinuum H1-1 trapped ion 20-qubit device as well as the code to run them via Quantinuum's API are provided in the \href{https://arxiv.org/src/2303.02209}{arXiv Tex Source folder}.

 \section*{Data availability}
 The binary data obtained from a Quantinuum H1-1 trapped ion 20-qubit device and shown in Fig.~\ref{fig: results} is provided in the \href{https://arxiv.org/src/2303.02209}{arXiv Tex Source folder}.

\bibliography{references}

\begin{acknowledgments}
We thank Fr\'ed\'eric Sauvage, Martin Eckstein and Cinthia Huerta Alderete for helpful conversations.
This work  used resources of the Erlangen National High Performance Computing Center and the Oak Ridge Leadership Computing Facility, which is a DOE Office of Science User Facility supported under Contract DE-AC05-00OR22725. 
 This work was supported by the German Research Foundation (DFG) – Project-ID 429529648 – TRR 306 QuCoLiMa, the German Federal Ministry of Education and Research (BMBF) contract number 13N16067 “EQUAHUMO”, and Munich Quantum Valley, which is supported by the Bavarian state government with funds from the Hightech Agenda Bayern Plus. 
 This work was supported by U.S. Department of Energy (DOE) via the Quantum Science Center (QSC), a National Quantum Information Science Research Center, and the Office of Science, Office of Advanced Scientific Computing Research, under the Accelerated
Research in Quantum Computing (ARQC) program. 
 The research for this publication was supported by a grant from the Priority Research Area DigiWorld under the Strategic Programme Excellence Initiative at Jagiellonian University.
TE acknowledges support from the International Max-Planck Research School for Physics of Light.
PC acknowledges support by the National Science Centre (NCN), Poland under project 2019/35/B/ST3/01028.
Work at Los Alamos was carried out under the auspices of the U.S. Department of Energy (DOE) National Nuclear Security Administration (NNSA) under Contract No. 89233218CNA000001, and was supported by the Laboratory Directed Research and Development program of Los Alamos National Laboratory under project number 20220253ER (JXZ, ATS), as well as by 20230049DR~(LC). ZH acknowledges initial support from the LANL Mark Kac Fellowship and subsequent support from the Sandoz Family Foundation-Monique de Meuron program for Academic Promotion. 
\end{acknowledgments}

\section*{Author contributions}
T.E. developed the quantum algorithm with support from Z.H., A.S., L.C. and P.C.. T.E. ran all classical numerics and implemented the quantum experiment on Quantinuum's API. T.E. and R.M. derived the algorithm error bounds and estimates; supported by Z.H. and M.J.H.. J.Z. proposed the physical system for our case study. L.C. numerically derived the ground state initialization circuits; with contributions from T.E.. All authors discussed and contributed to writing the manuscript.

%
%
\ \\

\onecolumngrid

\appendix

\section*{Appendix}
 Here, we first provide for pedagogical purposes a detailed derivation of the kick operator and effective Hamiltonian in Section~\ref{app:derivation} that was originally derived in Refs~\cite{goldman2014periodically, goldman2015erratum, eckardt2015high}. We then in Section~\ref{app:cases} spell out the specific form of the kick operator and effective Hamiltonian in a number of commonly encountered cases. 
In section~\ref{appendix: numerics} we discuss three different models as use cases for QHiFFS. These are a BNNNI model driven by a transverse magnetic field and a BNNNI model driven by a 2-local Pauli term, as well as a driven XY model whose effective dynamics need to be trotterized.
In the following section~\ref{appendix: kick_error}, we derive error bounds for the QHiFFS algorithms and provide a general estimate of the QHiFFS error as well as a tight estimation of the QHiFFS BNNNI error, that is shown in Fig.~\ref{fig:error_scaling} of the main text. In the last section~\ref{appendix: trotter_error} we give as a comparison error estimates and bounds for the QHiFFS competitor, standard Trotterization.

\renewcommand{\thesection}{Appendix \Alph{section}}
\section{Derivation of kick operator and effective Hamiltonian}\label{app:derivation}
\renewcommand{\thesection}{\Alph{section}}
Here,  provide a detailed derivation of the effective Hamiltonian and kick operator following~\cite{goldman2014periodically, goldman2015erratum, eckardt2015high}. This is a more detailed version of the Algorithmic Methods section of that paper. The aim is to shift the time-dependence of the Hamiltonian $\hat H(t)$ into a time-dependent basis transformation in which the time evolution is generated by an effective, time-independent Hamiltonian, $\hat H_{\rm eff}$ (setting $\hbar = 1$),
\begin{align}
    &i \partial_t \ket{\psi} = \hat H(t) \ket{\psi} \implies i \partial_t \ket{\phi} = \hat H_{\rm eff} \ket{\phi(t)} \label{eq:SE}
    \qquad \textrm{for } \ket{\phi(t)} = e^{i \hat K(t)} \ket{\psi} \; .
\end{align}
where the kick operator ansatz $\hat K$ to generate the basis change, is a choice without loss of generality, as it needs to be a unitary transformation. Inserting the definition of $\ket{\phi}$ into the Schrödinger equation (Eq. \eqref{eq:SE}) gives a defining relation for the effective Hamiltonian 
\begin{align}
    \hat H_{\rm eff} = e^{i\hat K(t)} \hat H(t) e^{-i \hat K(t)} + i \left( \frac{\partial e^{i \hat K(t)}}{\partial t} \right) e^{-i \hat K(t)}. \label{appendix_eq:HiF_def}
\end{align}
The first term on the right-hand side of Eq. \eqref{appendix_eq:HiF_def} (= Eq.~\eqref{eq:HiF_def}) can be expanded using the Hadamard lemma, i.e. we have 
\begin{align}
    e^{i \hat K(t)} \hat H(t) e^{-i \hat K(t)} &= \sum_{q = 0}^\infty \frac{i^q}{q!} \comm{\hat K(t)_{(q)}}{\hat H(t)} \; ,
    \label{eq:Heff_Hadamard}
\end{align}
where we denote $q$-fold nested commutators by $\comm{\hat K(t)_{(q)}}{\hat H(t)} = \comm{\hat K(t)}{\comm{\hat K(t)_{(q-1)}}{\hat H(t)}}$. The second term can then be expanded using the identity $\frac{\partial}{\partial x} e^{\hat A(x)} = \int_0^1 e^{y \hat A(x)} \left( \frac{\partial}{\partial x} \hat A(x) \right) e^{(1-y) \hat A(x)} dy$ and the Hadamard lemma, to give 
\begin{align}
    i \left( \frac{\partial e^{i \hat K(t)}}{\partial t} \right) e^{-i \hat K(t)} &= - \int_0^1 e^{ix \hat K(t)} \left( \frac{\partial}{\partial t} \hat K(t) \right) e^{-ix\hat K(t)} dx = -\sum_{q=0}^\infty \frac{i^{q}}{(q+1)!} \comm{\hat K(t)_{(q)}}{\frac{\partial \hat K(t)}{\partial t}} \; .
    \label{eq:Kick_Hadamard}
\end{align}
Plugging Eq. \eqref{eq:Heff_Hadamard} and Eq. \eqref{eq:Kick_Hadamard} into Eq. \eqref{appendix_eq:HiF_def} gives then Eq.~\eqref{appendix_eq:HiF_def2} (= Eq.~\eqref{eq:HiF_def2}):
\begin{align}
    \hat H_{\rm eff} &= \sum_{q = 0}^\infty \frac{i^q}{q!} \comm{\hat K(t)_{(q)}}{\hat H(t)}
    -\sum_{q=0}^\infty \frac{i^{q}}{(q+1)!} \comm{\hat K(t)_{(q)}}{\frac{\partial \hat K(t)}{\partial t}}  
    \label{appendix_eq:HiF_def2}
\end{align}
So far, this does not guarantee a solution beyond the trivial one where $\hat K(t) = \log( \mathcal{T} \exp(-i \int_{t_0}^t \hat H(s) ds))$ and $\exp(-i (t - t_0) \hat H_{\rm eff}) = \hat{ \mathds{1}}$. This is where the structure of the Hamiltonian, the periodic time-dependence comes in:
\begin{align}
    \hat H(t) = \hat H_0 + \hat V(t), \qquad \hat V(t) = \hat V(t+T).
\end{align}
Floquet's theorem then implies that the evolution operator can be transformed into Floquet normal form~\cite{floquet1883quations, holthaus2015floquet} where the dynamics is governed by a time-independent effective Hamiltonian and a time-dependent jitter term (which plays an analogous role to the kick operator). 

To obtain $\hat H_{\rm eff}$ and $\hat K(t)$ explicitly in terms of $\hat H_0$ and $\hat V(t)$, we use a high-frequency series expansion:
\begin{align}
		\hat H_{\rm eff} = \sum_{k=0}^\infty \frac{\hat H_{\rm eff}^{(k)}}{\omega^k} \; , \qquad\qquad \hat K(t) = \sum_{k=1}^{\infty} \frac{\hat K^{(k)}(t)}{\omega^k} \; .
		\label{eq:expansion}
\end{align}
with $\frac{T}{2\pi} = \frac{1}{\omega}$ and Fourier decomposition of the potential $\hat V(t) = \sum_{j=1}^\infty \hat V^{(j)} e^{i j \omega t} + \hat V^{(-j)} e^{-i j \omega t}$ to show its frequency dependence. Note that hermiticity of $\hat V(t)$ implies $\hat V^{(j)^\dagger} = \hat V^{(-j)}$. 
This gives the algorithm to obtain $\hat K(t)$ and $\hat H_{\rm eff}$ systematically order-by-order:
\RestyleAlgo{ruled}
\begin{algorithm}[hbt!]
\caption{for deriving $\tilde K(t)$ and $\tilde H_{\rm eff}$}
$k \gets 0$\;
\While{$k < k_{max}$}{
    \quad Consider Eq.~\eqref{appendix_eq:HiF_def2} up to order $k$\;
    \quad Plugin $\forall j \in \{1,...,k \}:\, \hat K^{(j)}(t)$ and $\hat H_{\rm eff}^{(j-1)}$ \;
    \quad $\hat K^{(k+1)}(t) \gets $ time-dependent terms\;
    \quad $\hat H_{\rm eff}^{(k)} \gets$ time-independent terms\;
    \quad $k \gets k + 1$\;
}
\end{algorithm} \ \\
Now, we demonstrate the method to find $\hat K(t)$ and $\hat H_{\rm eff}$ up to the second order. The terms contributing to $0^{th}$ order in Eq. \eqref{appendix_eq:HiF_def} are
\begin{align}
    \hat H_{\rm eff}^{(0)} = \hat H(t) - \frac{1}{\omega} \frac{\partial \hat K^{(1)}(t)}{\partial t} = \hat H_0 + \sum_{j=1}^\infty \left( \hat V^{(j)} e^{i j \omega t} + \hat V^{(-j)} e^{-i j \omega t} \right) - \frac{1}{\omega} \frac{\partial \hat K^{(1)}(t)}{\partial t} \; . \label{eq:Heff0}
\end{align}
Although $\hat K^{(1)}$ comes with a first order factor $\frac{1}{\omega}$, the term involving the time derivative $\frac{\partial}{\partial t} \hat K^{(1)} = \mathcal{O}(\omega)$ is of $0^{\rm th}$ order. This is because of the periodicity of $\hat K(t)$ and can be seen from the fact that the lowest frequency in its Fourier transform has to be $\omega$. To eliminate the time-dependence in Eq.~\eqref{eq:Heff0}, we define 
\begin{align}
    \hat K^{(1)}(t) = -i \sum_{j=1}^\infty \frac{1}{j} \left( \hat V^{(j)} e^{i j \omega t} - \hat V^{(-j)} e^{-i j \omega t} \right) \; ,
\end{align}
which leaves us with $\hat H_{\rm eff}^{(0)} = \hat H_0$. Next, $k=1$:
\begin{align}
    \frac{1}{\omega} \hat H_{\rm eff}^{(1)} &= \frac{i}{\omega} \comm{\hat K^{(1)}}{\hat H(t)} - \frac{i}{2\omega^2} \comm{\hat K^{(1)}}{\frac{\partial \hat K^{(1)}(t)}{\partial t}} - \frac{1}{\omega^2} \frac{\partial \hat K^{(2)}(t)}{\partial t} \\
    \iff \hat H_{\rm eff}^{(1)} &= \sum_{j=1}^\infty \frac{1}{j} \left( \comm{\hat V^{(j)}}{\hat H_0} e^{i j \omega t} + H.c. \right) + \frac{1}{2} \sum_{j,k=1}^\infty \frac{1}{j} \left( \comm{\hat V^{(j)}}{\hat V^{(k)}} e^{i (j+k) \omega t} + H.c. \right) \nonumber \\
    &+ \frac{1}{2} \sum_{j,k=1}^\infty \frac{1}{j} \left( \comm{\hat V^{(j)}}{\hat V^{(-k)}} e^{i (j-k) \omega t} + H.c. \right) - \frac{1}{\omega} \frac{\partial \hat K^{(2)}(t)}{\partial t} \; .
\end{align}
Again, we identify the time-dependent terms and integrate them to fix
\begin{align}
    \hat K^{(2)}(t) = &-i \sum_{j=1}^\infty \frac{1}{j^2} \left( \comm{\hat V^{(j)}}{\hat H_0} e^{i j \omega t} - H.c. \right) - \frac{i}{2} \sum_{j,k=1}^\infty \frac{1}{j(j+k)} \left( \comm{\hat V^{(j)}}{\hat V^{(k)}} e^{i (j+k) \omega t} - H.c. \right) \nonumber \\
    &- \frac{i}{2} \sum_{j\neq k=1}^\infty \frac{1}{j(j-k)} \left( \comm{\hat V^{(j)}}{\hat V^{(-k)}} e^{i (j-k) \omega t} - H.c. \right).
\end{align}
which gives $\hat H_{\rm eff}^{(1)} = \sum_{j=1}^\infty \frac{1}{j} \comm{\hat V^{(j)}}{\hat V^{(-j)}}$. Finally, we calculate $\hat H_{\rm eff}$ to second order,
\begin{align}
    \frac{1}{\omega^2} \hat H_{\rm eff}^{(2)} = &\frac{i}{\omega^2} \comm{\hat K^{(2)}}{\hat H(t)} - \frac{1}{2 \omega^2} \comm{\hat K^{(1)}}{\comm{\hat K^{(1)}}{\hat H(t)}} + \frac{1}{6 \omega^3} \comm{\hat K^{(1)}}{\comm{\hat K^{(1)}}{\frac{\partial \hat K^{(1)}(t)}{\partial t}}} \nonumber \\
    &- \frac{i}{2\omega^3} \comm{\hat K^{(1)}}{\frac{\partial \hat K^{(2)}(t)}{\partial t}} - \frac{i}{2\omega^3} \comm{\hat K^{(2)}}{\frac{\partial \hat K^{(1)}(t)}{\partial t}} - \frac{1}{\omega^3} \frac{\partial \hat K^{(3)}(t)}{\partial t} \; .
\end{align}
We can simplify this expression by first plugging in $\frac{\partial \hat K^{(1)}}{\partial t} = \omega \hat V(t)$. 
\begin{align}
    \hat H_{\rm eff}^{(2)} = &i \comm{\hat K^{(2)}}{\hat H_0} + \frac{i}{2} \comm{\hat K^{(2)}}{\hat V(t)} - \frac{1}{2} \comm{\hat K^{(1)}}{\comm{\hat K^{(1)}}{\hat H_0}} - \frac{1}{3} \comm{\hat K^{(1)}}{\comm{\hat K^{(1)}}{\hat V(t)}} \nonumber \\
    &- \frac{i}{2\omega} \comm{\hat K^{(1)}}{\frac{\partial \hat K^{(2)}(t)}{\partial t}} - \frac{1}{\omega} \frac{\partial \hat K^{(3)}(t)}{\partial t} \label{eq:Heff2}
\end{align}
As we do not intend to also derive the third order here, we drop all time-dependent terms to find,
\begin{align}
    \hat H_{\rm eff}^{(2)} = 
    &\frac{1}{2} \sum_{j=1}^\infty \frac{1}{j^2} \left( \comm{\comm{\hat V^{(j)}}{\hat H_0}}{\hat V^{(-j)}} + H.c. \right) + \frac{1}{4} \sum_{j,k=1}^\infty \frac{1}{j(j+k)} \left( \comm{\comm{\hat V^{(j)}}{\hat V^{(k)}}}{\hat V^{(-j-k)}} + H.c. \right) \nonumber \\
    &+ \frac{1}{4} \sum_{j\neq k=1}^\infty \frac{1}{j(j-k)} \left( \comm{\comm{\hat V^{(j)}}{\hat V^{(-k)}}}{\hat V^{(k-j)}} + H.c. \right) - \frac{1}{2} \sum_{j=1}^\infty \frac{1}{j^2} \left( \comm{\hat V^{(j)}}{\comm{\hat V^{(-j)}}{\hat H_0}} + H.c. \right) \nonumber \\
    &+\frac{1}{3} \sum_{j,k=1}^\infty \frac{1}{jk} \left( \comm{\hat V^{(j)}}{\comm{\hat V^{(k)}}{\hat V^{(-j-k)}}} + H.c. \right) - \frac{1}{3} \sum_{j \neq k=1}^\infty \frac{1}{jk} \left( \comm{\hat V^{(j)}}{\comm{\hat V^{(-k)}}{\hat V^{(k-j)}}} + H.c. \right) \nonumber \\
    &+\frac{1}{2} \sum_{j=1}^\infty \frac{1}{j^2} \left( \comm{\hat V^{(-j)}}{\comm{\hat V^{(j)}}{\hat H_0}} + H.c. \right) + \frac{1}{4} \sum_{j,k=1}^\infty \frac{1}{j(j+k)} \left( \comm{\hat V^{(-j-k)}}{\comm{\hat V^{(j)}}{\hat V^{(k)}}} + H.c. \right) \nonumber \\
    & + \frac{1}{4} \sum_{j\neq k=1}^\infty \frac{1}{j(j-k)} \left( \comm{\hat V^{(k-j)}}{\comm{\hat V^{(j)}}{\hat V^{(-k)}}} + H.c. \right) \\
    = &-\frac{1}{2} \sum_{j=1}^\infty \frac{1}{j^2} \left( \comm{\hat V^{(j)}}{ \comm{\hat V^{(-j)}}{\hat H_0}} + H.c. \right) +\frac{1}{3} \sum_{j,k=1}^\infty \frac{1}{jk} \left( \comm{\hat V^{(j)}}{\comm{\hat V^{(k)}}{\hat V^{(-j-k)}}} + H.c. \right) \nonumber \\
    &- \frac{1}{3} \sum_{j \neq k=1}^\infty \frac{1}{jk} \left( \comm{\hat V^{(j)}}{\comm{\hat V^{(-k)}}{\hat V^{(k-j)}}} + H.c. \right) \; ,
\end{align}
which has contributions from all terms of Eq. \eqref{eq:Heff2} except $i \comm{\hat K^{(2)}}{\hat H_0}$ and $- \frac{1}{\omega} \frac{\partial \hat K^{(3)}(t)}{\partial t}$, as these do not include time-independent terms. Putting together all terms up to second order, we find,
\begin{align}
    \tilde H_{\rm eff} = \hat H_0 &+ \frac{1}{\omega} \sum_{j=1}^\infty \frac{1}{j} \comm{\hat V^{(j)}}{\hat V^{(-j)}} - \frac{1}{2\omega^2} \sum_{j=1}^\infty \frac{1}{j^2} \left( \comm{\hat V^{(j)}}{ \comm{\hat V^{(-j)}}{\hat H_0}} + H.c. \right) \nonumber \\
    &+\frac{1}{3\omega^2} \sum_{j,k=1}^\infty \frac{1}{jk} \left( \comm{\hat V^{(j)}}{\comm{\hat V^{(k)}}{\hat V^{(-j-k)}}} + H.c. \right) - \frac{1}{3\omega^2} \sum_{j \neq k=1}^\infty \frac{1}{jk} \left( \comm{\hat V^{(j)}}{\comm{\hat V^{(-k)}}{\hat V^{(k-j)}}} + H.c. \right) + \mathcal{O}\left( \frac{1}{\omega^3} \right) \label{eq: Heff_series_2nd_order} \; , \\
    \tilde K(t) = &-\frac{i}{\omega} \sum_{j=1}^\infty \frac{1}{j} \left( \hat V^{(j)} e^{i j \omega t} - \hat V^{(-j)} e^{-i j \omega t} \right) - \frac{i}{\omega^2} \sum_{j=1}^\infty \frac{1}{j^2} \left( \comm{\hat V^{(j)}}{\hat H_0} e^{i j \omega t} - H.c. \right) \nonumber \\
    &- \frac{i}{2 \omega^2} \sum_{j,k=1}^\infty \frac{1}{j(j+k)} \left( \comm{\hat V^{(j)}}{\hat V^{(k)}} e^{i (j+k) \omega t} - H.c. \right) \nonumber \\
    &- \frac{i}{2\omega^2} \sum_{j\neq k=1}^\infty \frac{1}{j(j-k)} \left( \comm{\hat V^{(j)}}{\hat V^{(-k)}} e^{i (j-k) \omega t} - H.c. \right) + \mathcal{O}\left( \frac{1}{\omega^3} \right)
    \label{eq: Kick_series_2nd_order}
\end{align}
where $\tilde .$ denotes the high-frequency truncated forms of $\hat K(t)$ and $\hat H_{\rm eff}$. 
It becomes apparent from the above derivation that all orders of $\hat K(t)$ and $\hat H_{\rm eff}$ consist of commutator terms which later simplifies the calculation of the 2-norm. Let us note that in this work, the Fourier components of the driving potential commute, which further simplifies the high-frequency expansion.

\renewcommand{\thesection}{Appendix \Alph{section}}
\section{Special cases}\label{app:cases}
\renewcommand{\thesection}{\Alph{section}}
\begin{table}[b!]
\begin{tabularx}{\linewidth}{
    p{30mm} |
    p{32.5mm} |
    X |
    >{\centering\arraybackslash} p{40mm}
}
{Property of $\hat H_0$} & {Property of $\hat V(t)$} & {Implications and implementation strategy} & {Form of $\hat H_{\rm eff}$ and $\hat K(t)$} \\ \hline \hline
\multirow{4}{*}{$[\hat H_0, \hat V(t)] = 0 \, \forall t$} & 
\multirow{2}{*}{Single frequency} &
Trivial case. Standard Trotterization and the kick operator ansatz coincide. & \multirow{2}{*}{(\ref{eq:Heff_triv} - \ref{eq:Kick_triv})} \\ \cmidrule{2-4}
 & 
 \multirow{2}{*}{Multiple frequencies} & 
 Different Fourier components $\hat V^{(j)}$ might not commute which creates effective dynamics. & 
 \multirow{2}{*}{(\ref{eq:Heff_comm} - \ref{eq:Kick_comm})} \\ \hline
\multirow{5}{*}{$[\hat H_0, \hat V(t)] \neq 0$} & 
\multirow{3}{*}{Single frequency} & 
Corrections only from $[\hat H_0, \hat V(t)]$. This allows us to re-use previous time-independent techniques and lift them to Floquet Hamiltonian simulation. & 
\multirow{3}{*}{(\ref{eq:Heff_sw} - \ref{eq:Kick_sw})} \\ \cmidrule{2-4}
 & 
 \multirow{2}{*}{Multiple frequencies} & 
 Most general case taking into account all effective dynamics. & 
 \multirow{2}{*}{(\ref{eq: Heff_series_2nd_order} - \ref{eq: Kick_series_2nd_order})} 
\end{tabularx}
\caption{Comparison of special cases for the application to kick approximation}
\label{tab:special_cases}
\end{table}
Table \ref{tab:special_cases} lists different special cases that can simplify expressions for $\tilde H_{\rm eff}$ and $\tilde K(t)$. We discuss the four classes of periodic systems in the following. Starting with the expressions Eq.~\eqref{eq: Heff_series_2nd_order} and Eq.~\eqref{eq: Kick_series_2nd_order} derived above, assume the periodic potential only admits a single frequency 
 $\hat V(t) = \cos(\omega t + \phi_0) \hat W$, with $\hat W$ being time-independent. The Fourier representation then reads $\hat V(t) = \hat V^{(1)} e^{i ( \omega t+ \phi_0)} + \hat V^{(-1)} e^{-i ( \omega t+ \phi_0)} = \hat W/2 \left( e^{i ( \omega t+ \phi_0)} + e^{-i ( \omega t+ \phi_0)}\right)$. From $\hat V^{(1)} = \hat V^{(-1)} = \hat W/2$ follows $\comm{\hat V^{(1)}}{\hat V^{(-1)}}=0$ and we obtain: 
\begin{align}
    \hat H_{\rm eff} &= \hat H_0 - \frac{1}{2\omega^2} \left( \comm{\hat V^{(1)}}{ \comm{\hat V^{(-1)}}{\hat H_0}} + H.c. \right) + \mathcal{O}\left( \frac{1}{\omega^3} \right) \label{eq:Heff_sw} \; ,\\
    \hat K(t) &= -\frac{i}{\omega} \left( \hat V^{(1)} e^{i \omega t} - \hat V^{(-1)} e^{-i \omega t} \right) - \frac{i}{\omega^2} \left( \comm{\hat V^{(1)}}{\hat H_0} e^{i \omega t} - H.c. \right) + \mathcal{O}\left( \frac{1}{\omega^3} \right).
    \label{eq:Kick_sw}
\end{align}
Interestingly, the first order terms of $\hat H_{\rm eff}$ vanish. As long as the potential is 1-local, $\hat H_{\rm eff}$ will also not grow in locality, making it feasible to implement on quantum hardware without further compilation. Another simplification arises if $[\hat H_0, \hat V(t)] = 0 \, \forall t$ or equivalently $\comm{\hat H_0}{\hat V^{(j)}} = \comm{\hat H_0}{\hat V^{(j)}} = 0 \, \forall j$. If we admit arbitrary potentials, we obtain,
\begin{align}
    \hat H_{\rm eff} &= \hat H_0 + \frac{1}{\omega} \sum_{j=1}^\infty \frac{1}{j} \comm{\hat V^{(j)}}{\hat V^{(-j)}} - \frac{1}{3 \omega^2} \sum_{j,k=1}^\infty \frac{1}{j(j+k)} \left( \comm{\hat V^{(j+k)}}{\comm{\hat V^{(-j)}}{\hat V^{(-k)}}} + H.c. \right) \nonumber \\
    &- \frac{1}{3\omega^2} \sum_{j \neq k=1}^\infty \frac{1}{j(j-k)} \left( \comm{\hat V^{(j-k)}}{\comm{\hat V^{(-j)}}{\hat V^{(k)}}} + H.c. \right) + \mathcal{O}\left( \frac{1}{\omega^3} \right) \label{eq:Heff_comm} \; ,\\
    K(t) &= -\frac{i}{\omega} \sum_{j=1}^\infty \frac{1}{j} \left( \hat V^{(j)} e^{i j \omega t} - \hat V^{(-j)} e^{-i j \omega t} \right) - \frac{i}{2 \omega^2} \sum_{j,k=1}^\infty \frac{1}{j(j+k)} \left( \comm{\hat V^{(j)}}{\hat V^{(k)}} e^{i (j+k) \omega t} - H.c. \right) \nonumber \\
    &- \frac{i}{2\omega^2} \sum_{j\neq k=1}^\infty \frac{1}{j(j-k)} \left( \comm{\hat V^{(j)}}{\hat V^{(-k)}} e^{i (j-k) \omega t} - H.c. \right) + \mathcal{O}\left( \frac{1}{\omega^3} \right) \; .
    \label{eq:Kick_comm}
\end{align}
Here, $\hat H_0$ does not appear in higher order terms. This makes quantum simulation particularly simple, if the potential is 1-local since then $\hat K(t)$ and all corrections in $\hat H_{\rm eff}$ will be 1-local. For $p$-local $\hat V$, locality grows in general which introduces additional challenges for implementation. If both simplifications, $\comm{\hat H_0}{\hat V(t)}=0$ and a single frequency driving with $\comm{\hat V^{(j)}}{\hat V^{(-j)}} = 0$, are given, the lowest order kick approximation becomes exact
\begin{align}
    \hat H_{\rm eff} &= \hat H_0  \label{eq:Heff_triv} \\
    \hat K(t) &= -\frac{i}{\omega} \left( \hat V^{(1)} e^{i \omega t} - \hat V^{(-1)} e^{-i \omega t} \right). \label{eq:Kick_triv}
\end{align}
Since $\hat K(t)$ just denotes an integration of $\hat V(t)$, this case reduces to a simple case of Trotterization.

In the third case of Table \ref{tab:special_cases}, namely $\comm{\hat H_0}{\hat V(t)} \neq 0$ and a single frequency driving with $\comm{\hat V^{(1)}}{\hat V^{(-1)}} = 0$, known implementation techniques of the simulating $\hat H_0$ can be reduced or respectively lifted to Floquet Hamiltonian simulation. Since the first order correction in $\tilde H_{\rm eff}$ vanishes, simulating $\hat H_0$ already yields a good approximation to high-frequency dynamics. In the simplest non-trivial case, $\hat H_0$ is diagonal as in the BNNNI model discussed in the main text. Then, $\hat H_0$ can be simulated using a constant depth circuit. Similar cases of known implementations of $\hat H_0$ are 1D XY-models~\cite{verstraete2009quantum}, Bethe diagonalizable models \cite{sopena2022algebraic}, and translational invariant models \cite{mansuroglu2023variational}.

\renewcommand{\thesection}{Appendix \Alph{section}}
\section{Numerical Experiments}
\label{appendix: numerics}
\renewcommand{\thesection}{\Alph{section}}
In the main text, we provide numerical results for the same model as in our hardware demonstrations. Here, we provide further numerical examples to emphasize QHiFFS's capabilities in a broader setting. For comparability, we aim to keep the non-varying model parameters the same as in the main text; namely one-local coupling strength $h=2$, two-local nearest neighbor coupling strength $J_{(X,Y,Z)}=1$, next-nearest neighbor coupling $\kappa=0.25$, driving frequency $\omega = 30$ and system size $n_x\times n_y\, = \, 4 \times 5$.

In~\ref{appendix: numerics_BNNNI_X} we extend the numerical results on the periodically transversely driven BNNNI model, to the larger system sizes $n_x\times n_y\, = \, 4 \times 6$ and $n_x\times n_y\, = \, 5 \times 5$. In~\ref{appendix: numerics_BNNNI_XX}, we replace the one-local driving $\hat V(t) = - h  \cos(\omega t) \sum_i  \hat{X}_i $ with two-local nearest neighbor driving of the form $\hat V(t) = - J_X \cos(\omega t) \sum_{\langle i,j \rangle}  \hat{X}_i\hat{X}_j $. Further, we consider the influence of the next-nearest neighbor coupling by setting $\kappa=0$. And eventually in~\ref{appendix: numerics_XX_YY_cos_wt_ZZ}, we show that QHiFFS can even provide improvement for a periodically driven system, where the time evolution of the effective Hamiltonian $\hat H_{\rm eff}$ has to be trotterized.

\renewcommand{\thesection}{Appendix \Alph{section}}
\subsection{BNNNI model in a transverse field}
\label{appendix: numerics_BNNNI_X}
\renewcommand{\thesection}{\Alph{section}}
\begin{figure*}[t!]
    \centering
        \includegraphics[width=\linewidth]{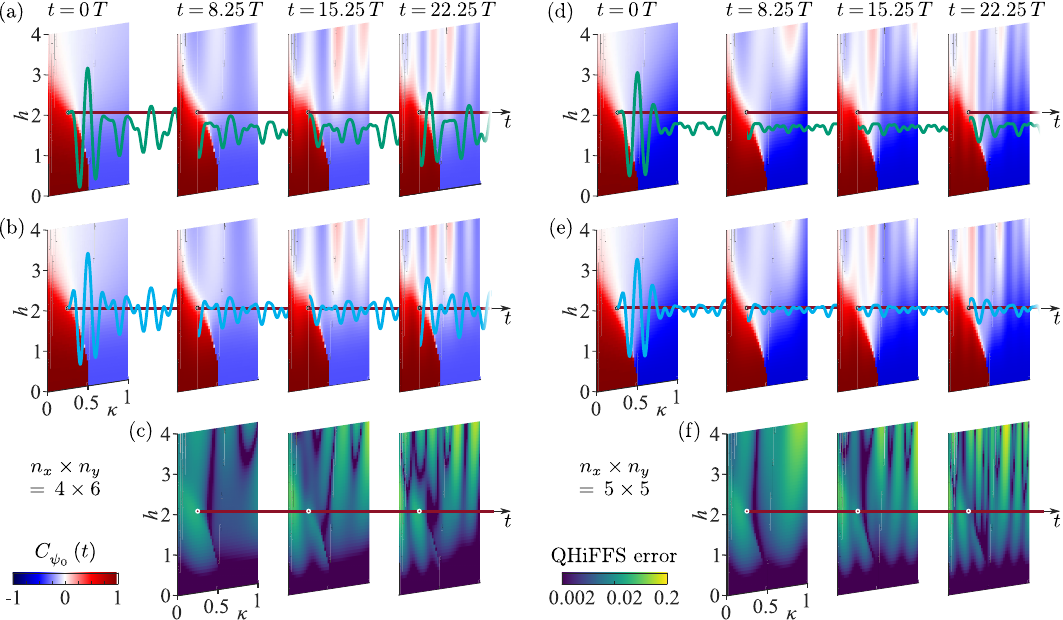}
        \vspace{- 0.5 cm}
    \caption{\textbf{Numerical simulations of non-equilibrium dynamics of the transverse field BNNNI model.}
    This figure is structured the same as Fig.~\ref{fig:ANNNI_numerics}
    in the main text. To extend the numerical results there, we show here in (a) - (c) corresponding results for $n_x\times n_y = 4 \times 6$ and in (d) - (f) for $n_x\times n_y = 5 \times 5$. The remaining model parameters are kept the same; namely $J=1$ and $\omega=30$.
    Accordingly, (a) and (d) show the next-nearest-neighbor correlation function, $C_{\psi_0}(t)$, as a function of $h$ and $\kappa$, where also the ranges are kept the same, for the same simulation times $t$ as in Fig.~\ref{fig:ANNNI_numerics} for exact time evolution. (b) and (e) give 
    $1^{\rm st}$ order QHiFFS time evolutions, that are feasible on current quantum hardware. The red line gives $C_{\psi_0}(t=0) \neq 0$ and the green/blue curves plot the exact/QHiFFS correlation function $C_{\psi_0}(t)$ as a function of time for the parameters $h=2$ and $\kappa=0.25$. 
    (c) and (f) depict the logscale error in the correlation function computed via QHiFFS. 
    Comparing $n_x\times n_y = 4 \times 5$, $4 \times 6$ and $5 \times 5$ yields no noticeable qualitative differences. In all cases, QHiFFS works similarly well, and similar features appear at similar times. The only noticeable difference is that for odd lattice lengths ($4 \times 5$ and $5 \times 5$) $C_{\psi_0}(t)$ takes slightly more negative values for $\kappa>1/2$ and $h$ small.  
    }
    \label{fig :BNNNNI_X_numerics_4x6_5x5}
\end{figure*}
To investigate the effect of system sizes and layouts, we here extend our numerical experiment from the main text, i.e. for the BNNNI model with one local transverse field driving to $n_x\times n_y = 4 \times 6$ and $n_x\times n_y = 5 \times 5$:
\begin{align}
    \hat{H}(t) = J \left( - \sum_{\langle i,j \rangle} \hat{Z}_i  \hat{Z}_j + \kappa \sum_{\langle\langle i,j \rangle\rangle}  \hat{Z}_i  \hat{Z}_j\right) - h  \cos(\omega t) \sum_i  \hat{X}_i \, ,
    \label{eq:H_BNNNI_sq}
\end{align}
This covers both slightly larger system sizes (up to $25 \%$) as well as the remaining even/odd combinations in $n_x$ and~$n_y$. 

Figure~\ref{fig :BNNNNI_X_numerics_4x6_5x5} (a) - (c) corresponds to Fig.~\ref{fig:ANNNI_numerics} of the main text for $n_x\times n_y = 4 \times 6$ and respectively (d) - (e) $n_x\times n_y = 5 \times 5$ instead of $n_x\times n_y = 4 \times 5$. All other features remain the same; namely the observable $C_{\psi_0}(t)$ (average next-nearest-neighbor correlation function, see main text Eq.~\eqref{eq:NNN}, $J=1$, $\kappa \in [0,1]$ $h \in [0,4]$, $\omega = 30$ and the final simulation times $t \,=\, 0 T, 8.25 T, 15.25 T, 22.25 T$. 

In comparison, all $n_x\times n_y = 4 \times 5$, $n_x\times n_y = 4 \times 6$ and $n_x\times n_y = 5 \times 5$ exhibit, as expected, widely similar features. For example, the transition regions of next-nearest neighbor correlated $C_{\psi_0}(t=0) > 0$, uncorrelated $C_{\psi_0}(t=0) \approx 0$ and anti-correlated $C_{\psi_0}(t=0) < 0$ ground states $\ket{\psi_0}$ are very similar in terms of $h$ and $\kappa$. In particular, the ferromagnetic regions ($C_{\psi_0}(t) > 0$, red regions in color plots) across system sizes $n_x\times n_y$ and final simulation times $t$ appear to be very comparable. 

Still, there are small differences. One of the main ones is that the minimal value of $C_{\psi_0}(t)$ in the anti-phase ($\kappa > 0.5$ and $J, \kappa \gg h$). For $n_x\times n_y = 4 \times 6$ it is less negative (lighter blue) than for $n_x\times n_y = 4 \times 5$ and $n_x\times n_y = 5 \times 5$. This is a well-understandable finite size effect originating also from the periodic boundary conditions. The ground states in the anti-phase are formed by symmetry equivalent superpositions of $2\times 2$ plaquettes. As the next-nearest neighbor correlation function measures the relative alignment of spins w.r.t to the $\hat Z$-direction and not their $\hat Z$ direction, for weak quantum fluctuations, i.e. small $h$, it is sufficient to consider only one plaquette state and only one column due to the periodic boundary conditions. Then one has for 5 qubits $\uparrow \uparrow \downarrow \downarrow \uparrow $ and for 6 qubits $\uparrow \uparrow \downarrow \downarrow \uparrow \uparrow $ (and symmetry equivalents). Consequently, all next-nearest neighbor correlations which cross the periodic boundary have value $+1$ and the ones that do not have value $-1$. This means for $h=0, \kappa > 1/2$ and $n_y=5$ that $C_{\psi_0}(t=0)=(-4+1)/5 = -3/5 $ and respectively for $n_y=6$ that $C_{\psi_0}(t=0)=(-4+2)/6 = -1/3$, which are the numerically observed values. Further, in the strongly driven regime $h>2$,  $n_x\times n_y = 4 \times 5$ and $n_x\times n_y = 4 \times 6$ are more alike than $n_x\times n_y = 5 \times 5$, which we may be related to the additional symmetry (protection), of the square lattice ground state. 

Importantly, QHIFFS works as well for $n_x\times n_y = 4 \times 6$ and $n_x\times n_y = 5 \times 5$ as for $n_x\times n_y = 4 \times 5$, which we ran on quantum hardware. For example, compare Fig.~\ref{fig :BNNNNI_X_numerics_4x6_5x5}(a) with (b) and (d) with (e) or respectively see (c) and (f). This high algorithmic accuracy indicates together with our $2\sigma$ sampling budget limited $n_x\times n_y = 4 \times 5$ hardware demonstration that indeed also $n_x\times n_y = 4 \times 6$ and $n_x\times n_y = 5 \times 5$ could have worked on quantum hardware.

\renewcommand{\thesection}{Appendix \Alph{section}}
\subsection{BNNNI Model with 2-local Driving Term}
\label{appendix: numerics_BNNNI_XX}
\renewcommand{\thesection}{\Alph{section}}
Here we explore using QHiFFS to simulate 2-local driving instead of 1-local driving. Specifically we consider,
\begin{align}
    \hat{H}(t) = J \left( - \sum_{\langle i,j \rangle} \hat{Z}_i  \hat{Z}_j + \kappa \sum_{\langle\langle i,j \rangle\rangle}  \hat{Z}_i  \hat{Z}_j\right) - J_X \cos(\omega t) \sum_{\langle i,j \rangle}  \hat{X}_i\hat{X}_j \, .
\end{align}
In this case, the non-locality of the effective Hamiltonian and the Kick operator grows faster (see Eq.~\eqref{eq: Heff_series_2nd_order} and Eq.~\eqref{eq: Kick_series_2nd_order}).
Fig.~\ref{fig: observables_BNNNI_ZZ_cos_wt_XX} shows two-point correlation functions similar to Fig.~\ref{fig: results}(a) and (b) of the main text.  
We additionally discuss the expected (on average) performance of QHiFFS using the average output fidelity as a function of final simulation time $t$, driving frequency $\omega$ and system size $n\,=\,n_x\times n_y$.

\subsubsection{Example simulation}
\begin{figure*}[t!]
    \centering
        \includegraphics[width=\linewidth]{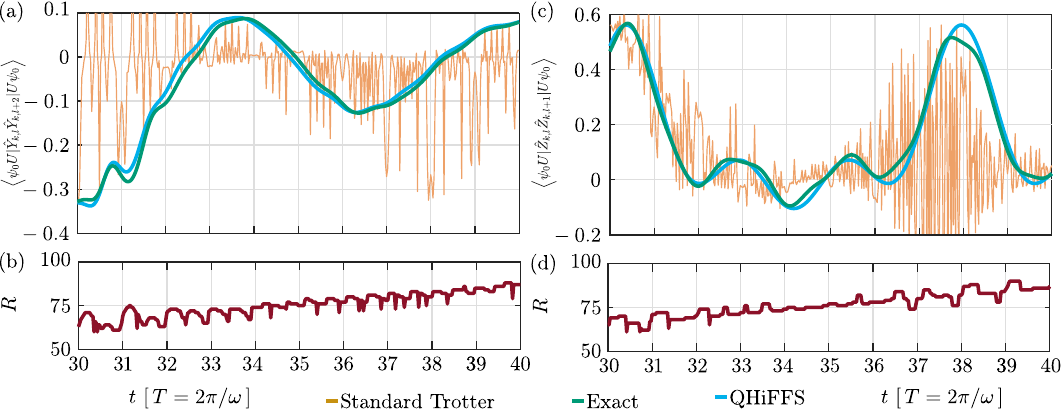}
    \vspace{- 0.5 cm}
    \caption{\textbf{Observable examples for the $\boldsymbol{\hat X_i \hat X_j}$-driven (BN)NNI model:} 
    This figure is organized equivalently to Fig.~\ref{fig: results}(a) and (b). 
    Here, (a) and (b) describe the time evolution of a $4 \times 4$ BNNNI model with nearest neighbor coupling $J=1$, nearest neighbor coupling $\kappa=0.25$, and two local driving $\hat V(t) = - cos(\omega t) \sum_{\langle i,j \rangle}  \hat{X}_i\hat{X}_j $ with driving frequency $\omega = 30$.  In (c) and (d) all parameters are the same apart from the next nearest neighbor coupling $\kappa$, which is set to 0, so that we re-obtain for the non-driven Hamiltonian $\hat H_0$ the nearest neighbor Ising model.
    In (a), we show the next-nearest-neighbor $\hat Y$-correlation function, as an observable example, for the time evolved ground state $\ket{\psi_0}$ and in (c) we show a nearest neighbor $\hat Z$-correlation function.
    In (a) and (c) the green line again gives the exact time evolution, 
    the blue line describes the time evolution by first-order QHiFFS ansatz, which again would be implementable on quantum hardware by a shallow circuit similar to Fig.~\ref{fig: quantinuum_circuit} in the main text, where the $\hat K(t)$-type circuit is replaced by a $\hat H_{\rm eff}$-type circuit, and the orange line gives same depth standard Trotterization. In (b) and (d) the red line again describes the improvement of first-order QHiFFS compared to standard second-order Trotter in terms of 2-qubit gate count $ R(t) = N_{\rm Trotter}(t)/N_{\rm QHiFFS}(t)$ to reach the same (noise-free) algorithmic fidelity. 
    }
    \label{fig: observables_BNNNI_ZZ_cos_wt_XX}
\end{figure*}

\begin{figure*}[t!]
    \centering
        \includegraphics[width=\linewidth]{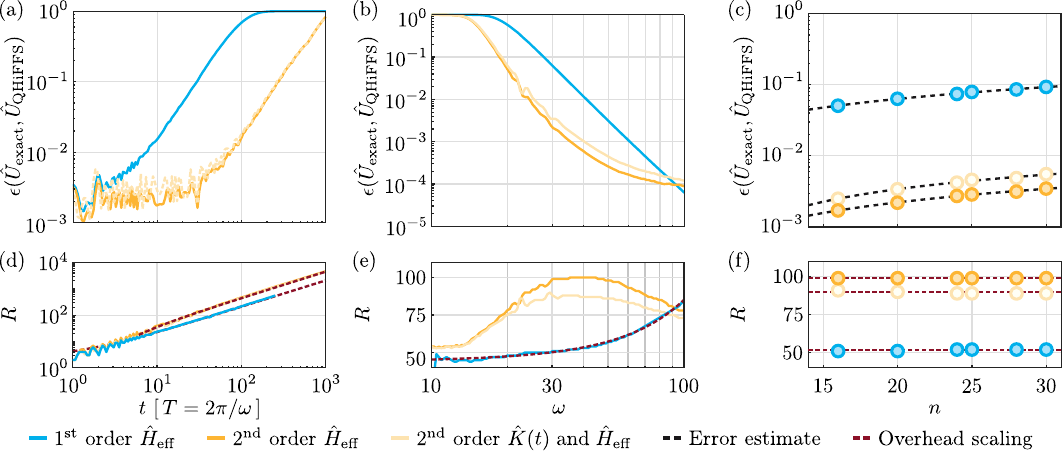}
\vspace{- 0.5 cm}
\caption{\textbf{Error analysis of the the $\boldsymbol{\hat X_i \hat X_j}$-driven BNNNI model.} 
    This figure is equivalent in structure, labeling, and coloring to Fig.~\ref{fig:error_scaling}, with the only difference that the transversal driving here is two local and given by $\hat V(t) = - \cos(\omega t) \sum_{\langle i,j \rangle}  \hat{X}_i\hat{X}_j $. All remaining model parameters are the same. Concretely, the non-driven time-independent part of the Hamiltonian $\hat H_0$ is as in the main text, with $J\,=\,1$, $\kappa\,=\,0.25$, $h\,=\,2$, $\omega\,=\,30$ and $t\,=\,22.75\,T$. Again, the columns show dependence on simulation time $t$, frequency $\omega$ as well as system size $n$, and the row gives the average algorithmic infidelity as well as the 2-qubit gate advantage $R$ of QHiFFS compared to standard Trotterization. In comparison with Fig.~\ref{fig:error_scaling}, there is a very good qualitative agreement but also quantitative similarities. This indicates that our in-depth error estimates of the one local driven BNNNI model, carry over to two local driving; namely $\epsilon(\hat U_{\textrm{exact}}, \hat U_{\textrm{QHiFFS}}) =\mathcal{O}(n t^2/ \omega^4)$ (see Eq.~\eqref{eq:error_ANNNI2D}) and $R=\mathcal{O}(t \omega^3)$ (see Eq.~\eqref{eq:error_comparision}. Those scaling dependencies can directly be read off the graphs (a) - (f). More generally, it also means that our derived upper error bounds in sections~\ref{app:Kick_upper_bound} and~\ref{appendix: trotter_error} are also tight here w.r.t to their scaling in $t$, $\omega$ and $n$.
    }
    \label{fig:error_scaling_ANNNI_cos_wt_XX}
\end{figure*}

In Fig.~\ref{fig: observables_BNNNI_ZZ_cos_wt_XX}(a) and (b), we show a time evolved next-nearest neighbor $\hat Y$ correlation function of the $4 \times 4$ BNNNI model with two-local nearest neighbor $\hat X$ driving. As in the main text, the nearest neighbor couplings $J=1$ and $J_X=1$, the next-nearest neighbor coupling $\kappa=0.25$, and the driving frequency $\omega=30$.  Respectively, (c) and (d) show a similar model, a nearest neighbor Ising model, which is the same apart from no next-nearest neighbor coupling ($\kappa = 0$). A first observation is that there is no significant performance difference of QHiFFS compared to the one local driving. First-order QHiFFS still can approximate the exact time evolution quantitatively very well (compare green and cyan lines). This is shown here even for later times than in the main text which is harder to simulate classically. Also equivalently and given more generally in  Fig.~\ref{fig:error_scaling_ANNNI_cos_wt_XX}(d) and Fig.~\ref{fig:error_scaling_ZZ_cos_wt_XX}(d), the circuit depth improvement $R$ for equivalent approximation fidelities in terms of 2-qubit gates still grows linearly. As before, at stroboscopic final simulation times $t\,=\, \frac{\mathbb{N}}{2}\,T$ we observe that standard Trotterization becomes relatively slightly better (compare small relative dips in Fig.~\ref{fig: observables_BNNNI_ZZ_cos_wt_XX}(b) and (d)), which coincides  with the kick operator term $\exp(-\hat K(t))$ approaching identity and simplifying  the QHiFFS approximation of the time evolution.

One of the main differences of two-local driving, compared to one-local driving is, that the average difference of two-local point correlation functions measured over time by the 1-norm are by a factor of 2$\frac{1}{4}$ to 3$\frac{1}{2}$ more larger (compare Fig.~\ref{fig: results} main text with Fig.~\ref{fig: observables_BNNNI_ZZ_cos_wt_XX}(a) and (c)). This is indeed beneficial, when one wants to reproduce and measure the shown time-dependent dynamics on shot-noise limited quantum hardware, as one would need a factor of 5 to 12 fewer measurements for the same relative accuracy. 

The circuit implementation of QHiFFS would still be very similar to the QHiFFS circuits of the one-local driven BNNNI model but with the one-local $\exp(-\tilde K(t))$ layer exchanged with a two-local one (comparable to $\exp(-i(t-t_0)\tilde H_{\rm eff})$).   The parameterized $\hat X_i \hat X_j$ gates can be transpiled into native ion trap parameterized $\hat Z_i \hat Z_j$ by a high fidelity layer of one local Hadamard gates before and after $\exp(-\tilde K(t))$ to realize the basis transformation between computational $ \hat Z$ basis and $\hat X$ basis. As before, the QHiFFS time evolution would still be approximately fast-forwardable. The two-qubit gate count for the $4\time 5$ BNNNI model would increase from 110 to 180 and respectively implementation of the $4\time 5$ NNI model would require 120 two-qubit gates. Considering our comparable 110 two-qubit gate shot-noise limited hardware implementation (see Fig.~\ref{fig: results}) and the lower sampling requirements, both implementations appear feasible.

\subsubsection{Algorithmic error behavior}
Having studied some or even many examples of time-evolved observables with respect to necessarily specific initial states naturally raises the question of whether these are particularly good, average, or even rather poor performance examples of the QHiFFS algorithm. To study the typical performance, we numerically calculate average algorithmic infidelities $\epsilon$ with respect to the Haar measure, which are both agnostic to the precise observables and initial states (compare Eq.~\eqref{eq:error_def}). 

\begin{figure*}[t!]
    \centering
        \includegraphics[width=\linewidth]{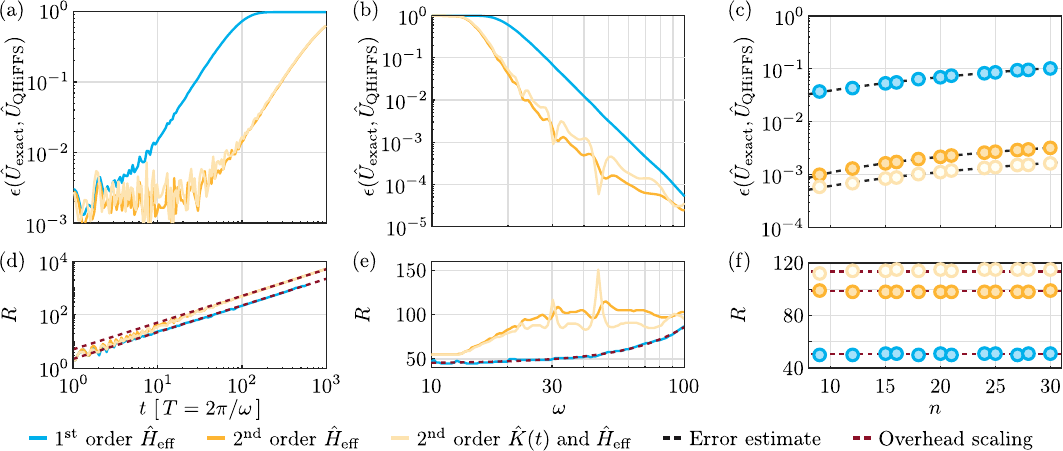}
\vspace{- 0.5 cm}
\caption{\textbf{Error analysis of the the $\boldsymbol{\hat X_i \hat X_j}$-driven 2D nearest neighbor Ising model.} 
    This figure corresponds to Fig.~\ref{fig:error_scaling} with $\hat V(t) = - \cos(\omega t) \sum_{\langle i,j \rangle}  \hat{X}_i\hat{X}_j $, $\kappa = 0$ and the remaining model parameters unchanged. There is a very good qualitative and quantitive agreement with Fig.~\ref{fig:error_scaling_ANNNI_cos_wt_XX}. One can read off (a) - (c) that the for one local driving predicted error scaling $\epsilon(\hat U_{\textrm{exact}}, \hat U_{\textrm{QHiFFS}}) =\mathcal{O}(n t^2 /\omega^4)$ (see Eq.~\eqref{eq:error_ANNNI2D}) still holds. Similarly, (d) - (f) show that the overhead scaling prediction $R=\mathcal{O}(t \omega^3)$ (see Eq.~\eqref{eq:error_comparision} also still holds. As before, those indicate that our derived error bounds are indeed asymptotically tight here (stressing their quality). 
    }
    \label{fig:error_scaling_ZZ_cos_wt_XX}
\end{figure*}

From Fig.~\ref{fig:error_scaling_ANNNI_cos_wt_XX}(a) and Fig.~\ref{fig:error_scaling_ZZ_cos_wt_XX}(a) we obtain algorithmic average infidelities $\epsilon(\hat U_{\textrm{exact}}, \hat U_{\textrm{QHiFFS}})$ of the order of 10 \% and a monotonous increase with time for $t\,\in \,[30\,T, 40\,T]$ shown in Fig.~\ref{fig: observables_BNNNI_ZZ_cos_wt_XX}(a) and (c) .
Both the initial state-specific two-qubit gate overheads $R$ in Fig.~\ref{fig: observables_BNNNI_ZZ_cos_wt_XX}(b) and Fig.~\ref{fig: observables_BNNNI_ZZ_cos_wt_XX}(d) as well as the initial state agnostic ones in  Fig.~\ref{fig:error_scaling_ANNNI_cos_wt_XX}(b), and ~\ref{fig:error_scaling_ZZ_cos_wt_XX}(b) for final simulations times $t\,\in \,[30\,T, 40\,T]$ are in the range of approx. 60 to approx. 90. This agreement between agnostic and specific errors as well as agnostic and specific overheads $R$ indicates that the examples shown in Fig.~\ref{fig: observables_BNNNI_ZZ_cos_wt_XX} are indeed typical cases.

For a quantum algorithm one of the main questions to address is how it scales with the number of qubits $n$.  We study how scaling advantages remain in the beyond classical regime, for both the BNNNI and NNI model with two-local transversal driving in  Fig.~\ref{fig:error_scaling_ANNNI_cos_wt_XX}(c), (f) and Fig.~\ref{fig:error_scaling_ZZ_cos_wt_XX}(c), (f). Analogously to the one local $\hat X$ driving in the main text, we find the average infidelity $\epsilon$ scales linearly with $n$ and that the improvement $R$ compared to standard Trotterization remains a constant factor. 

In summary, the errors here align with those of the one-local driving case$\epsilon(\hat U_{\textrm{exact}}, \hat U_{\textrm{QHiFFS}}) =\mathcal{O}(n t^2/ \omega^4)$ (see Eq.~\eqref{eq:error_ANNNI2D}) and $R=\mathcal{O}(t \omega^3)$ (see Eq.~\eqref{eq:error_comparision}). This large similarity between one and two local driving cases well justifies choosing the one-local drive scenario as we did in the main text for the hardware demonstration. 
Furthermore, the error scaling corresponds to the predicted dependencies by our general error bounds w.r.t to their scaling in $t$, $\omega$ and $n$ in section~\ref{app:Kick_upper_bound} and section~\ref{appendix: trotter_error}. This re-emphasises the quality of these general bounds. 

\renewcommand{\thesection}{Appendix \Alph{section}}
\subsection{XY Model with 2-local Driving Term}
\label{appendix: numerics_XX_YY_cos_wt_ZZ}
\renewcommand{\thesection}{\Alph{section}}
To conclude our numerical experiments, we discuss a case, where the truncated effective Hamiltonian $\tilde H_{\rm eff}$ has to be approximated by a time-independent Trotter formula. A periodically driven 2D XY model turns out to be suited for this endeavor. Consider the Hamiltonian
\begin{align}
    \hat{H}(t) = -J_X  \sum_{\langle i,j \rangle} \hat{X}_i  \hat{X}_j -J_Y  \sum_{\langle i,j \rangle} \hat{Y}_i  \hat{Y}_j  - J_Z \cos(\omega t) \sum_{\langle i,j \rangle}  \hat{Z}_i\hat{Z}_j.
\end{align}
The necessity of a product formula for $\tilde H_{\rm eff}$ creates a second source of error besides the QHiFFS error from the truncation of the high-frequency expansion, which we discuss in detail in Section~\ref{appendix: Numerics_XX_YY_cos_wt_ZZ_error_analysis} and lay out the theoretical background of in Section~\ref{appendix: trotter_error}.

\subsubsection{Example simulations}
\begin{figure*}[t!]
    \centering
        \includegraphics[width=\linewidth]{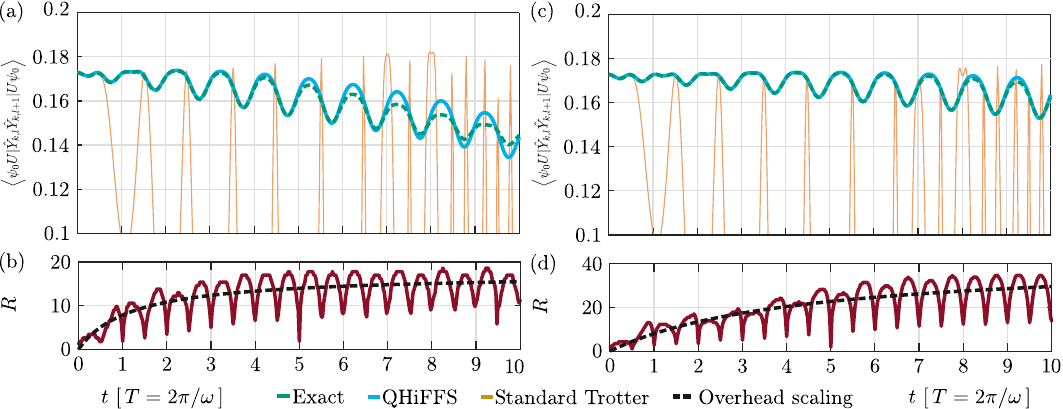}
    \vspace{- 0.5 cm}
    \caption{\textbf{Example simulations in the strong driving regime for the $\boldsymbol{\hat Z_i \hat Z_j}$-driven two dimensional XY model:} 
    This figure is structured the same as Fig.~\ref{fig: results} (a) and (b) in the main text as well as Fig.~\ref{fig: observables_BNNNI_ZZ_cos_wt_XX}. The exact time evolution is depicted in green, the QHiFFS approximately time evolution with one Trotter step ($m\,=\,1$) is shown in cyan, a Trotterization with the same circuit depth as QHiFFS is shown in orange and the predicted overhead scaling is in black. All subfigures are numerics on the $4 \times 4$ ${\hat Z_i \hat Z_j}$-driven two dimensional XY model with $J_Z\,=\,1$ and $J_X\,=\,J_Y$. In (a) and (b) $J_X\,=\, 0.1$ and in (c) and (d) $J_X\,=\, 0.05$.
    (a) and (c) plot the nearest neighbor $\hat Y$ two-point correlation functions, and (b) and (d) compares first-order QHiFFS with standard second-order Trotterization in terms of 2-qubit gate count to realize the same (noise-free) algorithmic fidelity $R(t) = N_{\rm Trotter}(t)/N_{\rm QHiFFS}(t)$. }
    \label{fig: observables_XX_YY_cos_wt_ZZ_small_Jxy}
\end{figure*}
\begin{figure*}[t!]
    \centering
        \includegraphics[width=\linewidth]{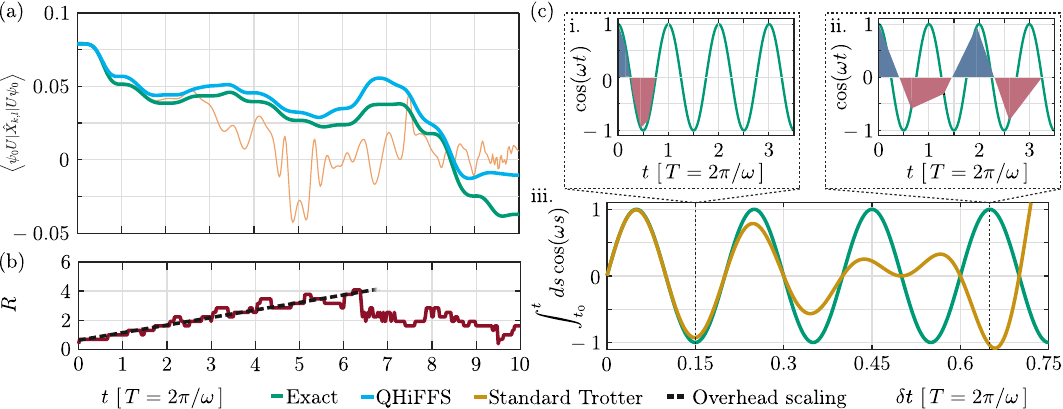}
\vspace{- 0.5 cm}
\caption{\textbf{Example simulations in a moderate-strength driving regime for the $\boldsymbol{\hat Z_i \hat Z_j}$-driven two dimensional XY~model:}  
(a) and (b) respectively show the time evolution of a correlation function and the overhead scaling compared to standard Trotterization. Again, green is the exact time evolution, cyan first order QHiFFS, orange same depth standard Trotterization, and black expected overhead scaling. In contrast to Fig.~\ref{fig: observables_XX_YY_cos_wt_ZZ_small_Jxy}, $J_X \,=\, J_Y \,=\, J_Z\,=\,1$ and five Trotter steps are used.
In (b) we see QHiFFS maintains a linear advantage over Trotterization at short times.
This indicates that here again the time-dependent discrete integration error again becomes dominant for standard Trotteriztation compared to the standard time-independent Trotterization error (compare also Eq.~\eqref{eq: appendix_std_Trotter_error}). 
(c) depicts an intuitive understanding of why and when standard (second-order) Trotterization breaks down. For this, we consider second-order numerical integration of $\cos (\omega s)$ (bottom) in orange and green being again the exact integration $\sin (\omega t)$. As in (a) and (b) we take five Trotter steps here. For small Trotter steps, e.g. $\delta t \, = \, t/m \, = \, 0.15\, T$ (first top panel), we see exact integration and numerical approximation matching well. Until $\delta t \, = \, T/2$ numerical and exact integration are still, in phase, despite the numerical value increasingly diverging; this means that the numerical value is still closer to the ground truth than random guessing. For $\delta t \, > \, T/2$, e.g. $=\, 0.65 \, T$ (right top panel), numerical integration becomes worse than random guessing and hence breaks down completely. 
Overall, this intuition is well matched by (a) and (b).
} 
    \label{fig: observables_XX_YY_cos_wt_ZZ_Jxyz_1}
\end{figure*}

Here, we show example simulations within two different regimes, the strong driving one, where $J_Z > J_X, J_Y$ (Fig.~\ref{fig: observables_XX_YY_cos_wt_ZZ_small_Jxy}), and the more moderate-strength one considered in the simulations in the main text where $J_Z \,=\, J_X \, = \, J_Y$ (Fig.~\ref{fig: observables_XX_YY_cos_wt_ZZ_Jxyz_1}). 
The approximate time evolutions that we compare here are given by:
\begin{align}
    \tilde U_{\rm QHiFFS}(t_0, t)  &= e^{-i\tilde K(t)} e^{-i(t - t_0)\tilde H_{\rm eff}}e^{+i\tilde K(t_0)} \, \nonumber \\
    &= \, 
    e^{-i F(t)\hat H_Z} 
    \left[ e^{i \frac{\delta t}{2} \hat H_X}
    \left(e^{i \delta t \hat H_Y} 
    e^{i \delta t \hat H_X} \right)^{m_Q}
    e^{-i \frac{\delta t}{2} \hat H_X } \right ]
    e^{+i F(t_0)\hat H_Z} 
    \label{eq: approx_QHiFFS_time_evolution_XX_YY_cos_wt_ZZ}
    \\
    \tilde U_{\rm Trotter}(t_0, t) \, 
    &= \, e^{i \frac{\delta t}{2} \hat H_X }
    \left(e^{i \frac{\delta t}{2} \hat H_Y } 
    e^{i \delta t f(t_0 + r \delta t) \hat H_Z }
    e^{i \frac{\delta t}{2} \hat H_Y}
    e^{i \delta t \hat H_X}
    \right)^{m_T}
    e^{-i \frac{\delta t}{2} \hat H_X}
    \label{eq: approx_std_Trotter_time_evolution_XX_YY_cos_wt_ZZ}
\end{align}
where second-order Trotterization is used both for $\tilde H_{\rm eff}$ and $\tilde H(t)$. The number of Trotter steps for QHiFFS and standard Trotterization are labeled with $m_Q$ and $m_T$ respectively. The step size is given by $\delta t = \frac{t_0 - t}{m}$. As in the main text, we focus here on first-order QHiFFS, as it results in shallower quantum circuits. The drivings, we consider here, are of the form $\hat V(t) = f(t) \hat H_Z$, resulting in kick operators of the form $\hat K(t) = F(t) \hat H_Z + \mathcal O (\frac{1}{\omega^2})$, with $F(t)$ being the antiderivative of $f(t)$. Taking $F(t_0) = 0$, we get from Eq.~\eqref{eq: approx_QHiFFS_time_evolution_XX_YY_cos_wt_ZZ} and Eq.~\eqref{eq: approx_std_Trotter_time_evolution_XX_YY_cos_wt_ZZ} that
\begin{align}
    R = \frac{4m_T + 2}{2m_Q + 3} \, ,
    \label{eq: R_XX_YY_cos_wt_ZZ}
\end{align}
compared to $R\,=\, m_T$ in the previously considered Ising-cases where the first-order $\tilde H_{\rm eff}$ is diagonal. 

One of the main principles of QHiFFS is to trade time-dependent errors, namely a time-ordering error, a discrete-integration as well as a commutator error of the type $\norm{\comm{\hat H_0}{\hat V(t)}}$ for a high-frequency one, while commutator errors within the time-independent truncated effective Hamiltonian $\tilde H_{\rm eff}$ cannot be avoided. 
Given that the high-frequency approximation error is not dominant, we can expect at least an improvement of the order of the avoided commutator error $\norm{\comm{H_0}{V(t)}}/ \norm{\comm{\hat H_X}{\hat H_Y}} \approx J_Z/J_Y+ J_Z/J_X$.

Fig.~\ref{fig: observables_XX_YY_cos_wt_ZZ_small_Jxy} shows results for the strong driving regime $J_Z > J_X, J_Y$, where we set $J_X \,= \, J_Y$ for simplicity. We find in this case that $R \rightarrow 2 J_X/J_Z$ as expected. As for the Ising-type models we also show the evolution of a $\hat Z_i\hat Z_j$-driven $4\times4$ nearest neighbor XY model with periodic boundary conditions, driving frequency $\omega \,=\,30$. In  (a) and (c) $J_X \,=\,0.1 $ and in (b) and (d) $J_X \,=\,0.05 $. As expected from Eq.~\eqref{eq: QHiFFS_norm_XX_YY_cos_wt_ZZ} and Eq.~\eqref{eq: Std_Trotter_norm_XX_YY_cos_wt_ZZ}, $R \, \approx \,  \frac{2 J_Z/J_X}{1+c/t}$ in (b) and (d), which is shown by the dashed black line.  

We additionally consider the driving regime where all couplings are the same strength, namely $J_X \,=\, J_Y \,=\, J_Z\,=\,1$. We take $m_Q \, = \,5$ to study multi-period behavior. We then expect from our error analysis (e.g. Eq.~\eqref{eq: QHiFFS_norm_XX_YY_cos_wt_ZZ}) that QHiFFS will work until $ t^3/m_Q^2 \, \lessapprox  \, 1$. This is indeed observed in Fig.~\ref{fig: observables_XX_YY_cos_wt_ZZ_Jxyz_1}(a), where the cyan QHiFFS time evolution follows closely the green time evolution until $t \, \approx \, 5 \, T$. In contrast standard Trotterization breaks already at $t/m_T \approx 1/2$. This is an indication for the dominance of the discrete integration error, specifically of $\cos (\omega t)$, which becomes worse than random for $\delta t \, > \, T/2$ (see Fig.~\ref{fig: observables_XX_YY_cos_wt_ZZ_Jxyz_1}(c)). Together, this explains why QHiFFS can still outperform standard Trotterization up to a factor of 4 (Fig.~\ref{fig: observables_XX_YY_cos_wt_ZZ_Jxyz_1}(b)); we get one factor 2 due to QHiFFS working twice as long for the the same number of Trotter steps $m$ and another factor of 2 as QHiFFS only needs 2 not 4 Hamiltonian exponential in each Trotter step (compare Eq.~\eqref{eq: approx_QHiFFS_time_evolution_XX_YY_cos_wt_ZZ} - Eq.~\eqref{eq: R_XX_YY_cos_wt_ZZ}).  Interestingly, in the equal coupling case, we regain the linear scaling of $R$ at short times (at longer times both QHiFFS and Trotterization break down). This observation is consistent with the high-frequency error term $\mathcal O (t/\omega^2)$ being dominant for QHiFFS and the time-independent Trotter and/or discrete integration error $\mathcal O (t^3/m_T^2)$ being dominant for standard Trotterization. 

\renewcommand{\thesection}{Appendix \Alph{section}}
\subsubsection{Algorithmic error behavior}
\label{appendix: Numerics_XX_YY_cos_wt_ZZ_error_analysis}
\renewcommand{\thesection}{\Alph{section}}
\begin{figure*}[t!]
    \centering
        \includegraphics[width=\linewidth]{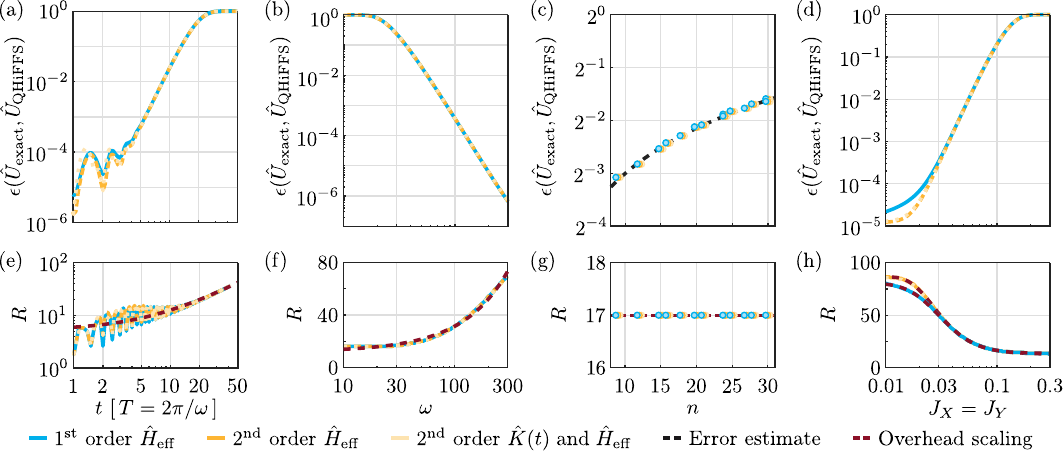}
\vspace{- 0.5 cm}
\caption{\textbf{Error analysis of the $\boldsymbol{\hat Z_i \hat Z_j}$-driven 2D nearest neighbor XY model.} 
Analogously to Fig.~\ref{fig:error_scaling} in the main text the first row shows the average infidelity of QHiFFS compared to exact time evolution and the second row gives the circuit depth advantage $R$ of QHiFFS, compared to standard Trotter in terms of 2-qubit gates.
The non-scanned parameters again align with our hardware implementation. Namely, we have $n_x \times n_y \, = \, 4 \times 5$ with periodic boundary conditions, time-independent coupling strength $J_X \, = \, J_Y \,=\, 0.1$, driving strength $J_Z \, = \, 1$, driving frequency $\omega \, = \, 30$, and final simulation time $t \, = \, 15.75 \, T $.
     }
    \label{fig:error_scaling_XX_YY_cos_wt_ZZ}
\end{figure*}

In contrast to the Ising-type models where QHiFFS was dominated by the high-frequency error and standard Trotterization was dominated by the discretized time integration errors, here both QHiFFS and standard Trotterization are no longer governed by a single dominant error source. Instead we have to deal with a mixture of several different error sources (compare Eq.~\eqref{eq: appendix_QHiFFS_error} and Eq.~\eqref{eq: appendix_std_Trotter_error}). Specifically, we can identify the following error sources as (compare Eq.~\eqref{eq:appendix_general_QHiFFS_error} and Eq.~\eqref{eq:appendix_general_Trotter_error}):
\begin{align}
    \epsilon \left( \hat U_{\rm exact}(t_0, t), \, \hat U_{\rm QHiFFS}(t_0, t) \right) &= \epsilon_{\rm high-frequency} + \epsilon_{\rm QHiFFS,\,time-indep. \, Trotter} 
    \label{eq: appendix_QHiFFS_error}
    \\
    \epsilon \left( \hat U_{\rm exact} (t_0, t), \,  \hat U_{\rm Trotter}(t_0, t) \right) &= \epsilon_{\rm time-ordering} + \epsilon_{\rm discrete \, integration} + \epsilon_{\rm  time-indep. \, Trotter}
    \label{eq: appendix_std_Trotter_error}
\end{align}
As noticed in the previous subsection, both QHiFFS and standard Trotterization will have a time-independent Trotter error contribution. When comparing these errors we get: 
\begin{align}
    \epsilon &\left( \hat U_{\rm exact} (t_0, t), \,  \hat U_{\rm Trotter}(t_0, t) \right) - \epsilon \left( \hat U_{\rm exact}(t_0, t), \, \hat U_{\rm QHiFFS}(t_0, t) \right) \nonumber \\
    &= \left( \epsilon_{\rm time-ordering} + \epsilon_{\rm discrete \, integration} + \epsilon_{\norm{\comm{H_0}{V(t)}}  \, \rm Trotter} \right )
    - \epsilon_{\rm high-frequency}  \, .
    \label{eq: appendix_std_Trotter-QHiFFS_error}
\end{align}
Hence, using the QHiFFS algorithm remains beneficial as long as the high-frequency error (negative sign) is smaller than the avoided Trotter errors (positive signs). 

Consistently with previous model examples, we again calculate in Fig.~\ref{fig:error_scaling_XX_YY_cos_wt_ZZ} numerically averaged infidelities for a $4\times5$ XY model driven by a nearest neighbor $\hat Z_i \hat Z_j$ coupling with $\omega \, = \, 30$ and coupling strength $J_Z \, = \, 1$. Further, we set $J_X \, = \, J_Y \,=\, 0.1$ and final simulation time $t \, = \, 15.75 \, T $, unless varied. A first difference to note compared to previous cases, is that there is no difference between first and higher-order frequency expansions (i.e., the cyan and yellow lines overlap). 
This suggests that indeed the high-frequency error (and thus the reason to prefer standard Trotterization compared to QHiFFS) is indeed non-dominant. This observation is further supported in Fig.~\ref{fig:error_scaling_XX_YY_cos_wt_ZZ}(a) and (d), where the dominant error scalings do not follow the high-frequency ones, but the time-independent Trotter error scaling of $\tilde H_{\rm eff}$ (compare also Eq.~\eqref{eq: Std_Trotter_norm_XX_YY_cos_wt_ZZ}). Also note here, that for $J_X, J_Y \rightarrow 0.01$ the $\tilde H_{\rm eff}$ Trotter error is decreased and as a consequence, the separation in high-frequency orders is reappearing. 

Our error analysis in Section~\ref{appendix: trotter_error} can match the more complex QHiFFS error scaling and also the standard Trotterization error scaling. This is why the numerically observed overhead $R$ (second row) fits very well to the overhead scaling we expect in dashed red. More specifically, for the final simulation time (Fig.~\ref{fig:error_scaling_XX_YY_cos_wt_ZZ}(e)) we see until $t\,\approx \, 10 \, T$, that $R$ ramps up to a constant advantage given by the avoided time-independent Trotter error and then regains a linear advantage, where typically the better in time scaling high-frequency error starts re-contributing to the overall error.
In terms of scaling advantage in driving frequency $\omega$, we see here only a linear scaling in Fig.~\ref{fig:error_scaling_XX_YY_cos_wt_ZZ}(f), not a cubic one. This is again due to the time-independent Trotter error of $\tilde H_{\rm eff}$ dominating the QHiFFS error ($\mathcal O (t^3/m_Q^2)$ in Eq.~\eqref{eq: QHiFFS_norm_XX_YY_cos_wt_ZZ}); while the standard Trotterization error is governed by the discretized integration error ($\mathcal O (\omega^2 t^3/m_T^2)$ in Eq.~\eqref{eq: Std_Trotter_norm_XX_YY_cos_wt_ZZ}).
As before, we see in Fig.~\ref{fig:error_scaling_XX_YY_cos_wt_ZZ}(g), that the scaling advantage is maintained for larger system sizes, even for this more complex example. This is explained well by a simple counting argument of the error terms, as those increase linearly with system size both for QHiFFS and standard Trotterization. Overall the error sources here are much more intertwined. This can result in the QHiFFS advantage being given by the avoided time-independent Trotter error of QHiFFS compared to standard Trotterization ($\norm{\comm{H_0}{V(t)}}$). Importantly, we then compare time-independent errors rather than the time-dependent ones. QHiFFS predominantly (but not exclusively) improves the time-dependent errors. This is why we selected a Ising-type model for our hardware demonstration in the main text.

\renewcommand{\thesection}{Appendix \Alph{section}}
\section{QHiFFS Error Analysis}
\label{appendix: kick_error}
\renewcommand{\thesection}{\Alph{section}}
In order to understand the error of the QHiFFS algorithm, we study the truncated tail of the high frequency expansion in this section. For a $k^\text{th}$ order QHiFFS algorithm, we calculate the leading error term when truncating the high frequency expansion in \ref{app:lead} and apply our results to two of the toy models discussed in this paper, the BNNNI model in \ref{app:BNNNI} and the XY model with 2-local driving in \ref{app:XY}. In \ref{subsec: Dependence on dimension and locality}, we further give special emphasis to the coefficient for the leading time dependent error term that scales as $\mathcal{O}(t^2)$ and depends on lattice dimension and locality of the interaction.

\renewcommand{\thesection}{Appendix \Alph{section}}
\subsection{Leading error terms}
\renewcommand{\thesection}{\Alph{section}}
\label{app:lead}
    Closely related to the output fidelity of a quantum simulation is the Hilbert-Schmidt product of its unitaries (cf. Eq.~\eqref{eq:error_def}), which we will focus on in the following error analysis. The error $\xi$ which we derive in the following appears in the Hilbert-Schmidt product as follows $\Tr\left( \hat U_1^\dagger \hat U_2 \right) = d(1 - \xi)$ with the Hilbert space dimension~$d$.  In general, we have
    \begin{align}
        \epsilon(\hat U_1,\hat U_2) = 1 - \frac{1}{d^2} 
        \Tr\left( \hat U_1^\dagger \hat U_2 \right)^2 =: 1 - (1 - \xi)^2 = 2 \xi + \mathcal{O}(\xi^2).
    \end{align}
    $\xi$ can thus be seen as an estimate of the infidelity of a Haar random state undergoing evolution by $\hat U_1$ and $\hat U_2$ up to a factor of 2. Using permutation invariance of the trace, the Hilbert-Schmidt product between a first order QHiFFS ansatz and exact evolution reads
\begin{align}
	\Tr(\hat U_{\rm exact}^\dagger (t_0, t) \hat U_{\rm QHiFFS}(t_0, t)) = \Tr\left( e^{i \tilde K (t_0)} e^{-i\hat K(t_0)} e^{i (t - t_0) \hat H_{\rm eff}} e^{i\hat K (t)} e^{-i \tilde K(t)} e^{-i (t - t_0) \tilde H_{\rm eff}} \right), \label{eq:error}
\end{align}
    where $\tilde K (t)$ denotes the truncated Kick operator and $\hat K (t)$ contains all orders (similar for $\hat H_{\rm eff}$). As the trace of a commutator is always zero, $\Tr\left(\comm{\hat A}{\hat B}\right) = 0$, we will omit terms that contribute with a vanishing commutator term. We want to expand Eq. \eqref{eq:error} step-by-step in orders of $\frac{1}{\omega}$ starting with the exponentials of $\hat K (t)$ and $\tilde K (t)$. Using the Baker-Campbell-Hausdorff (BCH) lemma
\begin{align}
	e^{i \hat K (t)} e^{-i \tilde K (t)} = &\hat{ \mathds{1}} + i \frac{1}{\omega^2} \hat K^{(2)}(t) - \frac{1}{2\omega^4} \left(\hat K^{(2)}(t)\right)^2 + \mathcal{O}\left(\frac{1}{\omega^5}\right).
\end{align}
Since the kick operator, $\hat K(t)$, as well as the effective Hamiltonian, $\hat H_{\rm eff}$, have a leading order error quadratic in $\frac{1}{\omega}$, which is a pure commutator, we expect the leading error of Eq. \eqref{eq:error} to be of fourth order as only products of commutators survive in the trace. The remaining factor will then be a similar combination of the exponentials of the effective Hamiltonians. This can be calculated with the following identity that follows from the Hadamard-Lemma
\begin{align}
    e^{\hat A} e^{-\hat A+ \hat \chi} - \hat{ \mathds{1}} = \int_{0}^{1} ds \, e^{s\hat A} \hat \chi \, e^{-s\hat A} + \mathcal{O}(\hat \chi^2) = \sum_{q=0}^{\infty} \frac{1}{(q+1)!} \comm{\hat A_{(q)}}{\hat \chi} + \mathcal{O}(\hat \chi^2) \; , \label{eq:Lemma}
\end{align}
as before we denoted $q$-fold nested commutators by $\comm{\hat A_{(q)}}{\hat \chi} = \comm{\hat A}{\comm{\hat A_{(q-1)}}{\hat \chi}}$. Applied to the remaining factors of Eq. \eqref{eq:error}, we define the second order error term of the effective Hamiltonian $\hat H_{\rm eff}$ as
\begin{align}
	e^{i(t - t_0) \hat H_{\rm eff}} e^{-i(t - t_0) \tilde H_{\rm eff}} &= \hat{ \mathds{1}} + \sum_{q=0}^{\infty} \frac{i^{q+1}(t - t_0)^{q+1}}{(q+1)!} \frac{1}{\omega^2} \comm{ \hat H_{0(q)}}{\hat H_{\rm eff}^{(2)}} + \mathcal{O}\left(\frac{1}{\omega^3}\right)\nonumber \\
    &=: e^{\Delta_{\rm eff} (t-t_0)} = \hat{ \mathds{1}} + \frac{1}{\omega^2} \hat \Delta_{\rm eff}^{(2)}(t - t_0)   + \mathcal{O}\left(\frac{1}{\omega^3}\right), \label{eq:Deff}
\end{align} 
where we used $\hat A:= i (t - t_0)\hat H_{\rm eff}$ and $\hat \chi = -i (t - t_0) (\hat H_{\rm eff} - \tilde H_{\rm eff})$. $\hat \Delta_{\rm eff}(t)$ again consists of commutators only and will consist of a series of infinitely many non-vanishing terms, in general. Later, we will use the integral formula \eqref{eq:Lemma} to evaluate $\hat \Delta_{\rm eff}^{(2)}(t)$ for the model under consideration. Taking all factors together and plugging back into Eq. \eqref{eq:error}, we obtain
\begin{align}
	\Tr(\hat U_{\rm exact}^\dagger(t_0, t) \hat U_{\rm QHiFFS}(t_0, t)) = d &+ \frac{1}{\omega^4} \Tr\Bigg( \frac{1}{2} \left(\hat \Delta_{\rm eff}^{(2)}(t - t_0)\right)^2 - i \hat K^{(2)} (t_0) \hat \Delta_{\rm eff}^{(2)}(t - t_0) - i \hat K^{(2)} (t) \hat \Delta_{\rm eff}^{(2)}(t_0 - t) + \nonumber \\
	&+ \hat K^{(2)}(t_0) e^{i (t - t_0) \hat H_0} \hat K^{(2)}(t) e^{-i (t - t_0) \hat H_0} - \frac{1}{2} \left( \hat K^{(2)}(t_0) \right)^2 - \frac{1}{2} \left( \hat K^{(2)}(t) \right)^2 \Bigg)
    + \mathcal{O}\left(\frac{1}{\omega^5}\right)\; , \label{eq:error_O4}
\end{align}
with $d = \Tr(\hat{ \mathds{1}})$ being the dimension of the Hilbert space. Note that, while $\hat \Delta_{\rm eff}(t)$ is in general linear in the simulation time, $t$, the errors coming from the kick operator are constant amplitude oscillations. One can generalize Eq. \eqref{eq:error_O4} to higher-order kick approximations; the general error term of a $(k-1)^{\rm st}$ order approximation reads
\begin{align}
	\Tr(\hat U_{\rm exact}^\dagger(t_0, t) \hat U_{\rm QHiFFS}(t_0, t)) = d &+ \frac{1}{\omega^{2k}} \Tr\Bigg( \frac{1}{2} \left( \hat \Delta_{\rm eff}^{(k)}(t - t_0)\right)^2 - i \hat K^{(k)} (t_0) \hat \Delta_{\rm eff}^{(k)}(t - t_0) \nonumber \\
 &- i \hat K^{(k)} (t) \hat \Delta_{\rm eff}^{(k)}(t - t_0)  
    + \hat K^{(k)}(t_0) e^{it \hat H_0} \hat K^{(k)}(t) e^{-it \hat H_0} \nonumber \\
    &- \frac{1}{2} \left(\hat K^{(k)}(t_0)\right)^2 - \frac{1}{2} \left(\hat K^{(k)}(t)\right)^2 \Bigg) + \mathcal{O}\left(\frac{1}{\omega^{2k+1}}\right). \label{eq:error_Ok}
\end{align}
With this general analysis, we find that the leading error term for high frequencies grows as
\begin{align}
    \Tr(\hat U_{\rm exact}^\dagger(t_0, t) \hat U_{\rm QHiFFS}(t_0, t)) - d = \mathcal{O} \left( \frac{(t - t_0)^2}{\omega^{2k}} \alpha_{\rm comm}^{(k)} \right).\label{eq:scaling1}
\end{align}
$\alpha_{\rm comm}^{(k)}$ includes norms of commutator terms that contain problem specific parameters. We provide a rigorous proof for this scaling in Sec.~\ref{app:Kick_upper_bound}.

\renewcommand{\thesection}{Appendix \Alph{section}}
\subsection{Application to BNNNI}
\label{app:BNNNI}
\renewcommand{\thesection}{\Alph{section}}
Here, we will discuss the error of the approximated evolution of the BNNNI model in two dimensions that is introduced in Eq.~\eqref{eq:error_def} of the main text for $t_0 = 0$ more in detail. The second order contributions of $\hat K (t)$ and $\hat H_{\rm eff}$ are
    \begin{align}
        \hat H_{\rm eff}^{(2)} &= 2Jh^2 \left( \sum_{\langle i,j \rangle} \left( \hat Z_i \hat Z_{j} - \hat Y_i \hat Y_{j} \right) + \kappa \sum_{\langle\langle i,j \rangle\rangle} \left( \hat Z_i \hat Z_{j} - \hat Y_i \hat Y_{j} \right) \right) + \mathcal{O}\left( \frac{1}{\omega^3}\right) \; ,\\
		\hat K^{(2)} (t) &= - 2Jh \cos(\omega t) \left( \sum_{\langle i,j \rangle} \hat Y_i \hat Z_{j} - \kappa \sum_{\langle\langle i,j \rangle\rangle} \hat Y_i \hat Z_{j} \right) + \mathcal{O}\left( \frac{1}{\omega^3}\right) \; .
    \end{align}
    To calculate the terms involving $\hat \Delta_{\rm eff}(t)^{(2)}$, we can apply the identities \eqref{eq:Lemma} and \eqref{eq:Deff}. We will show the derivation of the quadratic term. Other terms involving $\hat \Delta_{\rm eff}(t)^{(2)}$ vanish due to a mismatch of Pauli strings that are traceless. First, we use permutation invariance and change of integration variables to reduce the number of exponents in the expression
    \begin{align}
        \frac{1}{2} \Tr\left( \left(\hat \Delta_{\rm eff}^{(2)} (t) \right)^2 \right) = -\frac{1}{2} \Tr\left( \int_0^t ds \int_{-s}^{t-s} du \, e^{-iu\hat H_0} \hat H_{\rm eff}^{(2)} e^{iu\hat H_0} \hat H_{\rm eff}^{(2)} \right) \; .
        \label{eq:TFIM2D_E1}
    \end{align}
    Next, we simplify $e^{-iu\hat H_0} \hat H_{\rm eff}^{(2)} e^{iu\hat H_0}$ under the integral. In the end, only Pauli strings that appear twice, once in $e^{-iu\hat H_0} \hat H_{\rm eff}^{(2)} e^{iu\hat H_0}$ and once in $\hat H_{\rm eff}^{(2)}$ will have a non-vanishing trace. Considering this fact, we will gather irrelevant terms in the term $\hat \zeta$. 
    \begin{align}
        e^{-iu\hat H_0} \hat H_{\rm eff}^{(2)} e^{iu\hat H_0} = 2Jh^2 \Bigg ( &\sum_{\langle i,j \rangle} \hat Z_i \hat Z_j + \kappa^2 \sum_{\langle\langle i,j \rangle\rangle} \hat Z_i \hat Z_j  \nonumber \\
        &-  \cos^6(2Ju) \cos^6(2J\kappa u) \left ( \sum_{\langle i,j \rangle} \hat Y_i \hat Y_j +\kappa^2\sum_{\langle\langle i,j \rangle\rangle} \hat Y_i \hat Y_j \right ) \Bigg ) + \hat \zeta \; ,
        \label{eq:BNNNI_trick}
    \end{align}
    While the terms $\hat Z_i \hat Z_j$ commute with $\hat H_0$ and therefore stay unchanged, the $\hat Y_i \hat Y_j$ terms will collect factors from every non-commuting term in $\hat H_0$. The $\cos$ factors are derived using the identity for the exponential of a Pauli string $\hat P$:~$e^{i\theta \hat P} = \cos\theta \hat{ \mathds{1}} + i \sin\theta \hat P$. Plugging back into Eq.~\eqref{eq:TFIM2D_E1}, we get a dominant quadratic term $\frac{281}{64} J^2 h^2 (1+\kappa^2) t^2$ followed by a series of oscillatory terms that do not scale in time. Plugging $\hat K^{(2)}(t)$ into the remaining terms of Eq.~\eqref{eq:error_O4} gives
    \begin{align}
        \frac{\omega^4}{n h^2} \left( 1 - \frac{1}{d} \Tr(\hat U_{\rm exact}^\dagger(t) \hat U_{\rm QHiFFS}(t)) \right) = &h^2 (1+\kappa^2) \Bigg( \frac{281}{64} J^2 t^2 + \frac{75 \sin^2(2Jt)}{512} + \frac{15\sin^2(4Jt)}{1024} + + \frac{\sin(6 J (1 + \kappa) t)^2}{18432 (1 + \kappa)^2} \nonumber \\
         &+ \frac{5\sin^2(6Jt)}{4608} + \frac{15\sin^2(2J(-3 + \kappa) t)}{2048(\kappa - 3)^2} + \frac{45 \sin^2(2J(-2 + \kappa) t)}{1024(\kappa - 2)^2} + \nonumber \\
         &+ \frac{225 \sin^2(2J(-1 + \kappa) t)}{2048(\kappa - 1)^2} + \frac{9 \sin(4 J (-1 + \kappa) t)^2)}{2048 (-1 + \kappa)^2} + \frac{\sin(6 J (-1 + \kappa) t)^2}{18432(-1 + \kappa)^2} + \nonumber \\
         &+ \frac{75 \sin(2 J \kappa t)^2}{512 \kappa^2} + \frac{15 \sin(4 J \kappa t)^2}{1024 \kappa^2} + \frac{5 \sin(6 J \kappa t)^2}{4608 \kappa^2} + \frac{225 \sin(2 J (1 + \kappa) t)^2}{2048 (1 + \kappa)^2} + \nonumber \\
         &+ \frac{9 \sin(4 J (1 + \kappa) t)^2}{2048 (1 + \kappa)^2} + \frac{45 \sin(2 J (2 + \kappa) t)^2}{1024 (2 + \kappa)^2} + \frac{15 \sin(2 J (3 + \kappa) t)^2}{2048 (3 + \kappa)^2} + \nonumber \\
         &+ \frac{3 \sin(2 J (-3 + 2 \kappa) t)^2}{1024 (3 - 2 \kappa)^2} + \frac{45 \sin(2 J (-1 + 2 \kappa) t)^2}{1024 (1 - 2 \kappa)^2} + \frac{45 \sin(2 J (1 + 2 \kappa) t)^2}{1024 (1 + 2 \kappa)^2} + \nonumber \\
         &+ \frac{3 \sin(2 J (3 + 2 \kappa) t)^2}{1024 (3 + 2 \kappa)^2} + \frac{3 \sin(2 J (-2 + 3 \kappa) t)^2}{1024 (2 - 3 \kappa)^2} + \frac{15 \sin(2 J (-1 + 3 \kappa) t)^2}{2048 (1 - 3 \kappa)^2} + \nonumber \\
         &+ \frac{15 \sin(2 J (1 + 3 \kappa) t)^2}{2048 (1 + 3 \kappa)^2 } + \frac{3 \sin(2 J (2 + 3 \kappa) t)^2}{1024 (2 + 3 \kappa)^2} \Bigg) + 8 J^2 (1 + \cos^2(\omega t)) (1+\kappa^2) \nonumber \\
        &- 16 J^2 \cos(\omega t) \cos^2(2Jt) \cos^4(2J\kappa t) \left( (1+\kappa^2) \cos^2(2Jt) - 3\sin(J\kappa t) \right)  \; .
        \label{eq:ANNNI_error}
    \end{align}
This expression for the QHiFFS error (using a 1$^{\rm st}$-order effective Hamiltonian and kick approximation) is exact to corrections in $\mathcal{O}\left(\frac{1}{\omega^6}\right)$. We plot it in full in Fig.~\ref{fig:error_scaling}(a) - (c) in the main text. At long times, the first term, which grows quadratically in $t$, dominates leading to the approximate scaling quoted in Eq.~\eqref{eq:error_comparision} in the main text, whereas the term $8J^2(1+\kappa^2)$ gives a time-independent minimal error. Note that Eq.~\eqref{eq:ANNNI_error} also encodes the error of an Ising model, which can be computed taking the limit $\kappa \to 0$. The linear factor $n$ above accounts for the number of interaction terms for our local Hamiltonian. In general, long range interactions can exhibit a maximal error scaling quadratically in $n$.

\renewcommand{\thesection}{Appendix \Alph{section}}
\subsection{Dependence on dimension and locality}
\label{subsec: Dependence on dimension and locality}
\renewcommand{\thesection}{\Alph{section}}
    Here we focus on the BNNNI model interaction type ($\hat Z \hat Z$) with periodic driving and alter spatial dimension, $D$, and locality of the interaction that will finally influence the number of mutually non-commuting Hamiltonians in $\hat H_{\rm eff}^{(2)}$, $2 n_n$. Most interesting for long times will be the quadratically scaling error term of the form
    \begin{align}\label{eq:ct2}
        \xi = c_{t^2} \frac{n J^2 h^4}{\omega^4} (1+\kappa^2) t^2 + o(t)
    \end{align} 
    The coefficient $c_{t^2}$ comes out of the integral in Eq.~\eqref{eq:TFIM2D_E1}. We want to generalize this expression in the following to arbitrary spatial dimensions $D$ that will leave the Hamiltonian the same but changes the number of nearest and next-nearest neighbor terms. The following derivation is straight-forwardly generalized to $p$-local $\hat Z^{\otimes p}$ interactions because $c_{t^2}$ only depends on the number of non-commuting terms for a fixed $\hat Y$-interaction in $H_{\rm eff}^{(2)}$ each contributing with a cosine factor in Eq.~\eqref{eq:TFIM2D_E1}. In this analysis, we assume the number of mutually non-commuting Hamiltonians to be even; as for an odd number of cosine factors, the integral vanishes. The coefficient of the dominant (linear) term of the inner integral of Eq. \eqref{eq:TFIM2D_E1} will then reduce to 
    \begin{align}
        c_{t^2} = 2D \left( 1+ \lim_{s \to \infty} \frac{1}{s} \int_0^s du \cos(2 u)^{2 n_n} \cos(2 \kappa u)^{2 n_n} \right),
        \label{eq:dim_coeff}
    \end{align}
    where the factor $D$ comes from the fact that the number of interaction terms scales linearly in the spatial dimension. The 1 in the brackets resembles the coefficient of the $\hat Z \hat Z$ interactions and the integral comes from the $\hat Y \hat Y$ interactions. The integral in Eq. \eqref{eq:dim_coeff} can be calculated assuming $\kappa \neq 1$ using a recursive formula. First define
    \begin{align}
        c_{a_1, a_2, n_1, n_2}^\kappa := \lim_{s \to \infty} \frac{1}{s} \int_0^s du \cos(2^{a_1} u)^{2 n_1} \cos(2^{a_2} \kappa u)^{2 n_2}.
    \end{align}
    Using the trigonometric identity $\cos^2 x = \frac{1}{2} \left( 1+ \cos(2x) \right)$ and the binomial theorem, we can write down relations between the coefficients
    \begin{align}
        c_{a_1, a_2, n_1, n_2}^\kappa := \frac{1}{2^{n_1 + n_2}} \sum_{x_1 = 0}^{\left \lfloor \frac{n_1}{2} \right \rfloor} \sum_{x_2 = 0}^{\left \lfloor \frac{n_2}{2} \right \rfloor} \binom{n_1}{2x_1} \binom{n_2}{2 x_2} c_{a_1+1, a_2+1, x_1, x_2}^\kappa.
        \label{eq:rec_coeff}
    \end{align}
    These can now be recursively calculated using the termination conditions $c_{a_1, a_2, 0, 0} = 1$ and $c_{a_1, a_2, 0, 1} = c_{a_1, a_2, 1, 0} = \frac{1}{2}$. Plugging in the data of a two-dimensional BNNNI model ($D=2, 2n_n=2n_1=2n_2=6$) into Eq. \eqref{eq:rec_coeff} indeed yields the coefficient $\frac{281}{64}$ that is consistent with Eq. \eqref{eq:ANNNI_error}. We have gathered a number of examples in Table \ref{tab:coeffs}.

    \bgroup
        \def\arraystretch{1.5}
    \begin{table}[t]
        \centering
        \begin{tabularx}{\linewidth}{p{12mm} | p{28mm} p{28mm} p{28mm} | p{28mm} p{28mm} p{28mm}}
            & \multicolumn{3}{c|}{\centering Next-nearest neighbor Ising (NNNI)} 
            & \multicolumn{3}{c}{\centering Nearest neighbor Ising (NNI)} \\ \hline \hline $D$ &  \centering 1  & \centering 2 & \centering 3 & \centering 1 & \centering 2 & \centering 3 \cr
            \hline $2 n_n$ & \centering  2 & \centering  6 & \centering  10 & \centering  2 & \centering  6 & \centering  10 \cr \hline
            $c_{t^2}$   & \centering 2.5   & \centering $\frac{281}{64} \approx 4.39$  & \centering $\frac{208515}{32768}\approx 6.36$ & \centering 3 & \centering 5.25 & $\frac{957}{128}\approx 7.48$
        \end{tabularx}
        \caption{List of error coefficients $c_{t^2} = 2D (1 + c_{1, 1, n_1, n_2}^\kappa)$, defined implicitly in Eq.~\eqref{eq:ct2}, applied to one-, two-, and three-dimensional ($D=1,2,3$) nearest neighbor Ising (NNI) and next-nearest neighbor Ising (NNNI) models. Note that the BNNNI model corresponds to a 2D NNNI model. $2n_n$ is the number of mutually non-commuting Hamiltonians in $\hat H_{\rm eff}^{(2)}$.}
        \label{tab:coeffs}
    \end{table}
    \egroup
    
    Note that in the limit of large $D$, the complicated integral contribution of Eq. \eqref{eq:dim_coeff} is vanishing making the dependence on $D$ asymptotically linear. Putting this together with Eq. \eqref{eq:scaling1}, we can write down the scaling of the dominant error term of a second-order kick approximation of the BNNNI model
    \begin{align}
        \Tr(\hat U_{\rm exact}^\dagger(t) \hat U_{\rm QHiFFS}(t)) - d = \mathcal{O} \left( \frac{t^2}{\omega^{4}} J^2 h^4 n D (1 + o(1)) \right),
        \label{eq:scaling2}
    \end{align}
    where $o(1)$ denotes a term that vanishes in the limit of large $D$.

\renewcommand{\thesection}{Appendix \Alph{section}}
\subsection{Application to the XY model with 2-local driving}
\label{app:XY}
\renewcommand{\thesection}{\Alph{section}}
We discuss error estimates for the XY model presented in~\ref{appendix: numerics}. This model will suffer from both QHiFFS error (cf.~\ref{app:BNNNI}) and Trotter errors with commutator scaling. The Hamiltonian reads
\begin{align}
    \hat{H}(t) = -J_X  \sum_{\langle i,j \rangle} \hat{X}_i  \hat{X}_j -J_Y  \sum_{\langle i,j \rangle} \hat{Y}_i  \hat{Y}_j  - J_Z \cos(\omega t) \sum_{\langle i,j \rangle}  \hat{Z}_i\hat{Z}_j \, .
    \label{eq:XYHam}
\end{align}
Effective Hamiltonian and Kick operator read
\begin{align}
    \hat H_{\rm eff} &= -J_X  \sum_{\langle i,j \rangle} \hat{X}_i  \hat{X}_j -J_Y  \sum_{\langle i,j \rangle} \hat{Y}_i \hat{Y}_j \nonumber \\
    &+ \frac{4}{\omega^2} J_Z^2 \sum_{\langle i,j \rangle} \left( J_X \hat{X}_i  \hat{X}_j + J_Y \hat{Y}_i  \hat{Y}_j \right) - \frac{4}{\omega^2} J_Z^2 \sum_{\langle i,j \rangle, \langle j,k \rangle, \langle k,l \rangle} \left( J_X \hat{Z}_i \hat{Y}_j  \hat{Y}_k \hat{Z}_l + J_Y \hat{Z}_i \hat{X}_j  \hat{X}_k \hat{Z}_l \right) + \mathcal{O}\left( \frac{1}{\omega^3} \right) \; ,\\
    \hat K(t) &= - \frac{2}{\omega} J_Z \sin(\omega t) \sum_{\langle i,j \rangle} \hat Z_i \hat Z_j + \frac{4}{\omega^2} \sin(\omega t) J_Z \sum_{\langle i,j \rangle, \langle j,k \rangle} \left( J_X \hat Z_i \hat Y_j \hat X_k - J_Y \hat Z_i \hat X_j \hat Y_k \right) + \mathcal{O}\left( \frac{1}{\omega^3} \right).
\end{align}
The effective Hamiltonian has to be decomposed into a Trotter sequence. The Trotter error contains the commutator terms and is estimated by
\begin{align}
    \frac{t^2}{2m_T \sqrt{2^n}} \sum_{i>j}\norm{\comm{H^{(i)}}{H^{(j)}}}_2 = \frac{t^2}{m_T} n J_X J_Y.
\end{align}
For an estimation of the QHiFFS error, we can no longer use the trick of Eq.~\eqref{eq:BNNNI_trick}, but an upper bound that we derive in the following section.

\renewcommand{\thesection}{Appendix \Alph{section}}
\subsection{QHiFFS Error bound}
    \label{app:Kick_upper_bound}
    \renewcommand{\thesection}{\Alph{section}}
    While the above analysis only calculates an estimate to the truncated tail, we show in the following a rigorous upper bound for the error of the QHiFFS algorithm. After fixing conditions for which the high frequency expansion is absolutely convergence in \ref{app:convergence}, we calculate a general upper bound considering QHiFFS and Trotter errors in \ref{app:bound} -- scaling exactly as the error estimate derived in Eq.~\eqref{eq:error_Ok} -- and apply it to the BNNNI model in \ref{app:local}. 

    \renewcommand{\thesection}{Appendix \Alph{section}}
    \subsubsection{Absolute convergence of Kick operator and effective Hamiltonian}
    \label{app:convergence}
    \renewcommand{\thesection}{\Alph{section}}
        Expanding $\hat K (t)$ and $\hat H_\textrm{eff}$ as suggested in Eq.~\eqref{eq:expansion} is well-defined only under certain conditions that we want to discuss here. To justify a perturbative treatment, the series in both, $\hat K(t)$ and $\hat H_{\textrm{eff}}$, need to be absolutely convergent. Let us define
        \begin{align}
            \norm{\hat K^{(j)}(t)} =: q_K^{(j)}(t)\ ,\qquad \qquad \norm{\hat H_\textrm{eff}^{(j)}} =: q_H^{(j)}(t).
        \end{align}
        The choice of the norm does not need to be specified at this point. We will later choose the 2-norm to compare to the error analysis of Sec.~\ref{appendix: kick_error}. Absolute convergence gives bounds to the series
        \begin{align}
            \sum_{j=1}^\infty \frac{q_{K}^{(j)}}{\omega^j} \leq \frac{C_K}{\omega} = \mathcal{O}\left( \frac{1}{\omega} \right) \ , \qquad \qquad \sum_{j=0}^\infty \frac{q_{H}^{(j)}}{\omega^j} \leq C_H = \mathcal{O}\left( 1 \right). \label{eq:series}
        \end{align}
        Note that absolute convergence is dependent on the value of $\omega$. The root test, for instance, gives a condition for which the series in Eq.~\eqref{eq:series} converges and also one for which it diverges. For
        \begin{align}
            \limsup_{j \to \infty}\sqrt[j]{\frac{q^{(j)}_{K/H}}{\omega^j}} = \limsup_{j \to \infty} \frac{\sqrt[j]{q^{(j)}_{K/H}}}{\omega} < 1.
            \label{eq:root_test}
        \end{align}
        the series is absolutely convergent. If we replace ``$<$'' by ``$>$'' it is a condition for divergence. In the equal case ``$=$'', no statement can be made, in general. As per construction, the coefficients $q_{K/H}^{(j)}$ do not depend on $\omega$, the above condition is always violated in the limit $\omega \to 0$. In the limit $\omega \to \infty$, the left-hand side of Eq.~\eqref{eq:root_test} tends towards 0 implying absolute convergence. Hence, there will always be a critical frequency $\omega_c$ that limits the regime in which a Kick approximation is well-defined or not. $\omega_c$ will depend on the energy scales of $\hat H_0$ and $\hat V(t)$, in general (for the BNNNI model, $\omega_c = \mathcal{O}(J\sqrt{1+\kappa^2},h)$). In the whole paper, we assume to be in the high-frequency regime $\omega > \omega_c$.
        
        Since Eq.~\eqref{eq:series} converges absolutely in the high-frequency regime, we can now write the remainders as
        \begin{align}
            &\sum_{j=k+1}^\infty \frac{q_{K}^{(j)}}{\omega^j} = \frac{\tilde C_K}{\omega^{k+1}} = \mathcal{O}\left( \frac{1}{\omega^{k+1}} \right) \ , \qquad \qquad \sum_{j=k+1}^\infty \frac{q_{H}^{(j)}}{\omega^j} = \frac{\tilde C_H}{\omega^{k+1}} = \mathcal{O}\left( \frac{1}{\omega^{k+1}} \right), \label{eq:error_remainder} \nonumber \\
            &\text{where we defined} \qquad \tilde C_{K/H} = \omega^{k+1} \lim_{N \to \infty} \sum_{j=k+1}^N \frac{q_{K/H}^{(j)}}{\omega^j}
        \end{align}
        The Kick approximation assumes a truncation of the series in $\hat K(t)$ and $\hat H_{\rm eff}$ at a specified order $k$. A generalized form of Eqs.~\eqref{eq: Heff_series_1st_order} and~\eqref{eq: Kick_series_1st_order} of the main text reads
        \begin{align}
            \tilde H_\textrm{eff} = \sum_{j=0}^k \frac{\hat H_\textrm{eff}^{(j)}}{\omega^j} \ , \qquad\qquad \tilde K(t) = \sum_{j=1}^{k} \frac{\hat K^{(j)}(t)}{\omega^j}.
        \end{align}
        The approximated time evolution is then constructed via
        \begin{align}
            \tilde U_{\rm QHiFFS} (t_0, t) = e^{-i \tilde K(t)} e^{-i(t-t_0) \tilde H_\textrm{eff}} e^{i \tilde K(t_0)}
            \label{eq:Kick_evol}
        \end{align}
        In the following, we derive an error bound for the difference of approximated and exact time evolution which is obtained by replacing the truncated series by the full series in Eq.~\eqref{eq:Kick_evol}. 

\renewcommand{\thesection}{Appendix \Alph{section}}
    \subsubsection{General bound}
    \label{app:bound}  
    \renewcommand{\thesection}{\Alph{section}}
        \begin{lemma}
            Let $\hat A_j, \hat B_j,\, j \in \{1, ..., n_j\}$ be lists of hermitian operators, then
            \begin{align}
                \norm{\prod_{j=1}^{n_j} e^{i \hat A_j} - \prod_{j=1}^{n_j} e^{i \hat B_j} } \leq \sum_{j=1}^{n_j} \norm{\hat A_j - \hat B_j},
            \end{align}
            where $\norm{.}$ is any norm that admits unitary invariance.
            \begin{proof}
                We begin with writing the difference operator $\prod_{j=1}^{n_j} e^{i \hat A_j} - \prod_{j=1}^{n_j} e^{i \hat B_j}$ in integral form
                \begin{align}
                    &\prod_{j=1}^{n_j} e^{i \hat A_j} - \prod_{j=1}^{n_j} e^{i\hat  B_j} = \int_0^1 dx \frac{d}{dx} \left[ \prod_{j=1}^{n_j} e^{i \left(\hat B_j + x(\hat A_j - \hat B_j) \right)} \right] \nonumber \\
                    &= \int_0^1 dx \int_0^1 dy \sum_{k=1}^{n_j} \Bigg[ \prod_{j=1}^{k-1} e^{i \left(\hat B_j + x(\hat A_j - \hat B_j) \right)} e^{i y \left(\hat B_k + x(\hat A_k - \hat B_k) \right)} (\hat A_k - \hat B_k) 
                    e^{i (1-y) \left(\hat B_k + x(\hat A_k - \hat B_k) \right)} \prod_{j=k+1}^{n_j} e^{i \left(\hat B_j + x(\hat A_j - \hat B_j) \right)} \Bigg].
                    \label{eq:diff_op}
                \end{align}
                In the first step, we used the fundamental theorem of calculus and in the second step, we used the product rule and again the integral identity $\frac{\partial}{\partial x} e^{\hat A(x)} = \int_0^1 e^{y \hat A(x)} \left( \frac{\partial}{\partial x} \hat A(x) \right) e^{(1-y) \hat A(x)} dy$. If we take the norm of Eq.~\eqref{eq:diff_op}, we can use the triangle inequality and its unitary invariance to get to the result
                \begin{align}
                    &\norm{\prod_{j=1}^{n_j} e^{i \hat A_j} - \prod_{j=1}^{n_j} e^{i \hat B_j}} \leq \nonumber \\&\leq \int_0^1 dx \int_0^1 dy \sum_{k=1}^{n_j}  \Bigg| \Bigg|\prod_{j=1}^{k-1} e^{i \left(\hat B_j + x(\hat A_j - \hat B_j) \right)} e^{i y \left(\hat B_k + x(\hat A_k - \hat B_k) \right)} (\hat A_k - \hat B_k) 
                    e^{i (1-y) \left(\hat B_k + x(\hat A_k - \hat B_k) \right)} \prod_{j=k+1}^{n_j} e^{i \left(\hat B_j + x(\hat A_j - \hat B_j) \right)} \Bigg|\Bigg| \nonumber \\
                    &=\int_0^1 dx \int_0^1 dy \sum_{j=1}^{n_j} \norm{\hat A_j - \hat B_j} 
                    =\sum_{j=1}^{n_j} \norm{\hat A_j - \hat B_j}.
                \end{align}
            \end{proof}
            \label{lem:Hamiltonian_bound}
        \end{lemma}
        Using Lemma \ref{lem:Hamiltonian_bound}, we can now bound the QHiFFS algorithm error
        \begin{align}
            \norm{\hat U_{\rm QHiFFS}(t_0, t) - \tilde U_{\rm QHiFFS}(t_0, t)} &\leq \norm{K(t) - \tilde K(t)} + |t-t_0| \norm{H_{\rm eff} - \tilde H_{\rm eff}} 
            + \norm{K(t_0) - \tilde K(t_0)} \nonumber \\
            &= \mathcal{O}\left( \frac{|t-t_0|}{\omega^{k+1}} + \frac{1}{\omega^{k+1}} \right),
            \label{eq:appendix_bound}
        \end{align}
        where the error scaling is a direct consequence of Eq.~\eqref{eq:error_remainder}. In the main text, we are comparing the norm squared, which then scales as $\mathcal{O}\left(\frac{|t-t_0|^2 + |t-t_0| + 1}{\omega^{2k+2}}\right)$.

        If $\tilde H_{\rm eff}$ is being trotterized, additional errors occur. In that case, we are not implementing $\tilde H_{\rm eff}$, but actually the BCH-Hamiltonian $H_{BCH}$ that governs the dynamics of the trotterized sequence. We can further bound 
        \begin{align}
            \norm{H_{\rm eff} - H_{BCH}} \leq \norm{H_{\rm eff} -  \tilde H_{\rm eff}} + \norm{ \tilde H_{\rm eff} - H_{BCH}} = \mathcal{O}\left( \frac{|t-t_0|}{\omega^{k+1}} + \frac{t^2}{m_T} \max_{i,j}\norm{\comm{H^{(i)}}{H^{(j)}}} \right),
            \label{eq: Heff_BCH}
        \end{align}
        where the first error is caused by the truncation of the high frequency series and the second term is a standard Trotter error of a time-independent Hamiltonian split into terms $\{H^{(j)}\}_j$. Note that the Trotter error is generally larger for time-dependent Hamiltonians (cf. \ref{appendix: trotter_error}), such that QHiFFS still remains favorable in cases where $H_{\rm eff}$ is not directly implementable on hardware.

\renewcommand{\thesection}{Appendix \Alph{section}}
    \subsubsection{Local interactions}
    \label{app:local}
    \renewcommand{\thesection}{\Alph{section}}
        In contrast to the QHiFFS error estimate derived in Sec.~\ref{appendix: kick_error}, the upper bound in Eq.~\eqref{eq:appendix_bound} does not resolve the dependency of the error on locality, spatial dimension and system parameters. If we specify the driving potential and reduce the discussion to 2-local interactions with finite interaction length in the non-driven Hamiltonian $\hat H_0$, we can refine Eq.~\eqref{eq:appendix_bound}. For instance, the BNNNI Hamiltonian from Eq.~\eqref{eq:BNNNI_Hamiltonian} of the main text has such a form. Since $\hat V(t)$ only acts on a single qubit, the locality of nested commutators between $\hat V^{(j)}$ and $\hat H_0$ does not grow. The 2-norms $q_{K/H}^{(j)}$ thus grow only with $\sqrt{n}$. The remainders $\tilde C$ defined in Eq.~\eqref{eq:error_remainder} read
        \begin{align}
            \frac{\tilde C_K }{\omega^{k+1}} &= \mathcal{O} \left( \frac{\sum_{j=0}^k J^j h^{k+1-j}}{\omega^{k+1}} \sqrt{n} \right) \ , \qquad \qquad \frac{\tilde C_H }{\omega^{k+1}} = \mathcal{O} \left( \frac{\sum_{j=0}^k J^j h^{k+2-j}}{\omega^{k+1}} \sqrt{n} \right).
        \end{align}

The time-dependent QHiFFS error estimate for the driven XY model (cf. Eq.~\eqref{eq:XYHam}) can be derived in a similar way
\begin{align}
    \frac{t}{\omega^2}\norm{H_{\rm eff}^{(2)}} = \frac{4t}{\omega^2} \sqrt{13n} J_Z \sqrt{J_X^2 + J_Y^2}.
\end{align}

\renewcommand{\thesection}{Appendix \Alph{section}}
\section{Standard Trotterization error estimate and bound}
\label{appendix: trotter_error}
\renewcommand{\thesection}{\Alph{section}}
    In general, even after rotating into the kicked frame, a Trotterization of the effective Hamiltonian is in order. Still, the Trotter error of the time-independent effective Hamiltonian is simpler and more robust to various system parameter scalings. Errors of time-dependent Hamiltonian simulation have been investigated before \cite{Berry2020timedependent, An2022timedependent}. To discuss this, consider a first order Trotter formula,
    
    \begin{align}
	   \hat U_{\rm Trotter}(t_0, t) = \prod_{r=1}^{m_T} \prod_{j=1}^{n_j} e^{i\delta t \hat H_j (t_0 + r\delta t)},
    \label{eq:TrotterO1}
    \end{align}
    where $\hat H(t) = \sum_{j=1}^{n_j} \hat H_j(t)$ is a choice of Hamiltonian decomposition and $\delta t = \frac{t - t_0}{m}$. Standard Trotter errors come from the non-commutativity of the Hamiltonians $\hat H_j(t)$. As the Hamiltonian of interest is time-dependent, there are two other sources of error coming from the non-commutativity of the Hamiltonian at different times and the discretization of the time-ordered integrals that is implicit in Eq. \eqref{eq:TrotterO1}. Using the triangle inequality, the error is bounded by
    \begin{align}
        \left| \left| \mathcal{T} e^{-i \int_{t_0}^{t} \hat H(s) ds }- \hat U_{\rm Trotter}(t_0, t) \right| \right| &\leq \left| \left| \hat U_{\rm exact}(t_0, t)  - e^{-i \int_{t_0}^{t} \hat H(s) ds } \right| \right| + \left| \left| \hat U_{\rm Magnus}(t_0, t)  - \prod_{r=1}^{m_T} e^{i\delta t \hat H(t_0 + r\delta t)} \right| \right| \nonumber \\
        \label{eq:appendix_Trotter_error_full}
        &+ \left| \left| \hat U_{\rm discrete}(t_0, t) - \prod_{r=1}^{m_T} \prod_{j=1}^{n_j} e^{i\delta t \hat H_j (t_0 + r\delta t)} \right| \right| 
        =: \xi_1 + \xi_2 + \xi_3,
    \end{align}
    The first term, $\xi_1$, in Eq. \eqref{eq:appendix_Trotter_error_full} describes errors from the non-commutativity of $\hat H(t)$ at different times scaling with $\max_{s, s'} \comm{\hat H(s)}{\hat H(s')}$, the second term, $\xi_2$, takes into account the discretization of the integral and the third term, $\xi_3$, is the standard Trotter error scaling with commutators $\comm{\hat H_i (s) }{\hat H_j(s)}$. $\xi_1 + \xi_3$ scales as $\mathcal{O}\left( \frac{t^2}{m_T} J(1+\kappa) h \right)$ for the BNNNI model. We will see that in the regime of large frequencies, $\omega$, the term $\xi_2$ will be dominant. Since, we are interested in the 2-norm squared, there will also be mixed terms taking into account $\xi_1, \xi_2, \xi_3$.

    For the sake of clarity we will now focus on $\xi_2$ only. To do so, we neglect the non-commutativity for a moment and consider just a time-discretization as described in Eq. \eqref{eq:V_disc}. If we neglect the non-commutativity of the Hamiltonian at different times $\comm{\hat H(s)}{\hat H(s')} = 0$, we can write the generator of discretized dynamics with the Heavyside step function $\Theta(x) = \begin{cases} 0 & x<0 \\ 1 & x \geq 0 \end{cases}$
    \begin{align}
        \left| \left| \hat U_{\rm Magnus}(t_0, t) - \hat U_{\rm discrete}(t_0, t) \right | \right |^2 &=  \left| \left| e^{-i \int_{t_0}^{t} \hat H(s) ds} - e^{-i \int_{t_0}^{t} \hat H_{\rm discrete}(s) ds}  \right | \right |^2 
        \label{eq:V_disc}
        \\
        \text{with } \hat H_{\rm discrete}(s) &= \sum_{r=1}^{m_T} \left( \Theta \left( s - (r-1) \delta t \right) - \Theta \left( s - r \delta t \right) \right) \hat H( r \delta t).
    \end{align}
    Using the inequality $||e^{i \hat H_1} - e^{i \hat H_2}|| \leq ||\hat H_1 - \hat H_2||$ (see Lemma 50 of Ref.~\cite{chakraborty2018power} for proof), we get:
    \begin{align}
		\left| \left| \hat U_{\rm Magnus}(t_0, t) - \hat U_{\rm discrete}(t_0, t) \right | \right |^2 \leq &\left|\left| \int_{t_0}^{t} ds \hat H(s) - \delta t \sum_{r=1}^{m_T} \hat H( t_0 + r \delta t)\right|\right|^2
        \label{eq:appendix_Trotter_bound}
	\end{align}
    The 2-norm can be related to the Hilbert-Schmidt product via $\frac{1}{2d} ||\hat U_1 - \hat U_2||^2 = 1 - \frac{1}{d} \Re (\Tr(\hat U_1^\dagger \hat U_2))$, so that the two measures can be directly compared, as $\Tr(\hat U_1^\dagger \hat U_2)$ is real by construction. 

    \enlargethispage{1 cm}
    For the BNNNI model, only the oscillating terms will survive in Eq. \eqref{eq:appendix_Trotter_bound}. As we are integrating over a cosine, integrations over every full period vanish. Coarse discretizations, however, will collect error terms. We divide the time interval $t - t_0 = l T + \delta$, $l \in \mathbb N$ into an interval of $l$ full periods and $\delta < T$. 
    \begin{align}
		&\left| \left| \hat U_{\rm Magnus}(t_0, t) - \hat U_{\rm discrete}(t_0, t) \right | \right |^2 \leq h^2 \left| \left| \sum_j \hat X_j \right|\right|^2 \left| \int_{t_0}^{t} ds \cos(\omega s) - \delta t \sum_{r=1}^{m_T} \cos(\omega (t_0 + r \delta t)) \right|^2 + \mathcal{O}(\delta t^2) \label{eq:hatX} \\
		\leq& h^2 n d \left | \int_{t - \delta}^{t} ds \cos(\omega s) - \delta t \sum_{r= m_T - m_\delta + 1}^{m_T} \cos(\omega (t_0 + r \delta t)) \right.  \left. - l \delta t \max_{l \in [0, ..., m_T-1]} \sum_{r = \frac{l}{k} (m_T - m_\delta) + 1}^{\frac{l+1}{l} (m_T - m_\delta)} \cos(\omega (t_0 + r \delta t)) \right |^2.
        \label{eq:Trot_error_interim}
	\end{align} 
    Here, we assume the number of Trotter steps, $m_T$, to be equally divided over the time interval which leaves $m_\delta$ steps for simulation $s \in [lT, lT+\delta ]$ and $m_T - m_\delta$ for the simulation of $l$ full periods. The last term of Eq. \eqref{eq:Trot_error_interim} describes the worst discretization of the $l$ period which yields an upper bound for every other period. Standard estimates of the Riemann sum errors give,
    \begin{align}
		\frac{1}{2d} &\left| \left| \hat U_{\rm exact}(t_0, t) - \hat U_{\rm discrete}(t_0, t) \right | \right |^2 \leq \frac{h^2 n}{2} \left( \frac{T^2 \omega m_T^2}{2({m_T} - m_\delta)} \max_{s\in[0,T]} |\sin(\omega s)| + \frac{\delta^2 \omega}{2m_\delta} \max_{s\in[0,\delta]} |\sin(\omega s)| \right)^2 \nonumber \\
		&\leq \frac{h^2 \pi^2 n}{2} \left( \frac{T {m_T}^2}{{m_T} - m_\delta} + \frac{\delta}{m_\delta} \right)^2 = \frac{h^2 \pi^2 n (t - t_0)^2}{2 {m_T}^2} \left( {m_T} + 1 \right)^2 = \frac{h^2 n (t - t_0)^2}{2 {m_T}^2} \left( \left(t - t_0 - \delta\right) \frac{\omega}{2} + \pi \right)^2 \; .
        \label{eq:appendix_Trotter_error}
	\end{align}
    In the last three steps, we further estimated $|\sin x| \leq 1$, $\delta \leq T$ and plugged in $\omega = \frac{2\pi}{T}$ and $\frac{m_\delta}{m_T} = \frac{\delta}{t - t_0}$, as well as $\frac{{m_T} - m_\delta}{l} = \frac{l T}{t - t_0}$ coming from the equal distribution of Trotter steps. Finally, we expressed the number of periods $l=(t - t_0 - \delta) \frac{\omega}{2 \pi}$ and the period, $T$, in terms of the driving frequency, $\omega$. In the special case $\delta = 0$, we can absorb the second term of Eq. \eqref{eq:appendix_Trotter_error} in $l$. The error then reads
    \begin{align}
        \frac{1}{2d} \left| \left| \hat U_{\rm Magnus}(t_0, t) - \hat U_{\rm discrete}(t_0, t) \right | \right |^2 &\leq \frac{h^2 n}{8 m_T^2} (t-t_0)^4 \omega^2.\label{eq:appendix_Trotter_error_delta0}
    \end{align}
    Note the quadratic dependence on $\omega$ that makes $\xi_2$ the dominant error in our setting. If both QHiFFS and Trotter approximations have comparable error, the Trotter number, $m_T$, can be bounded
    \begin{align}
        \eqref{eq:appendix_Trotter_error_delta0} \stackrel{!}{=} \eqref{eq:ANNNI_error} &\iff 
        \frac{h^2 n}{{m_T}^2} t^4 \omega^2 \geq \frac{281}{8} \frac{J^2 h^4 n}{\omega^4} (1+\kappa^2) t^2 + \mathcal{O}\left( \frac{1}{\omega^4} \right) + \mathcal{O}\left( \frac{t^2}{\omega^6} \right) \nonumber \\
        &\iff {m_T} \leq \frac{ t \omega^3}{J h \sqrt{1+ \kappa^2} + \mathcal{O}\left( \frac{1}{t} \right) + \mathcal{O}\left( \frac{1}{\omega} \right)} \nonumber \\
        &\implies {m_T} = \mathcal{O}\left( \frac{ t \omega^3}{J h \sqrt{1+ \kappa^2}} \right) = R.
        \label{eq:m_bound}
    \end{align}
    We have set $t_0 = 0$ for comparison without loss of generality. For the sake of simplicity, we have chosen the best case $\delta = 0$ for the Trotter sequence, so the second term of Eq. \eqref{eq:appendix_Trotter_error} is dropped. Also note that Eq. \eqref{eq:m_bound} only holds in the high-frequency regime in which Eq. \eqref{eq:ANNNI_error} is a good estimate and Eq. \eqref{eq:appendix_Trotter_error_delta0} is the dominant Trotter error. Finally, since the $1^{\rm st.}$ order QHiFFS requires the same number of two qubit gates as one ($1^{\rm st.}$~or $2^{\rm nd.}$ order) Trotter step, we have here that $m_T = R$ where $R$ is the ratio of the number of two-qubit gates required by standard Trotterization compared to QHiFFS to implement a simulation of the same fidelity. Note that the actual operator $\hat X_j$ in Eq.~\eqref{eq:hatX} can be exchanged by another Pauli without loss of generality. The discussion of this section hence straight-forwardly generalizes to other models considered in Sec.~\ref{appendix: numerics}.

\medskip
Let us summarize the error bounds for QHiFFS and standard Trotterization. QHiFFS involves a high frequency expansion that breaks down when the frequency gets too small $\omega \leq \norm{\hat V (t)}, \norm{\hat H_0}$, but also introduces errors from the Trotterization of the effective Hamiltonian with a Trotter number $m_Q$ and Trotter order~$q$, i.e. 
\begin{align}
    \norm{\hat U_{\rm exact} - \hat U _{\rm QHiFFS}} = \mathcal{O}\left( \frac{t}{\omega^{k+1}} \max_{(j_1, ..., j_k)}\norm{\comm{\hat V^{(j_1)}}{\comm{...}{\comm{\hat V^{(j_k)}}{\hat H_0}}...}} 
    +t \left( \frac{t}{m_Q}\right)^q \max_{i,...,j}\norm{\comm{\hat H_{\rm eff}^{(i)}}{\comm{...}{\hat H_{\rm eff}^{(j) } } } } \right) \, .
    \label{eq:appendix_general_QHiFFS_error}
\end{align}
The Trotter error, on the other hand, suffers from errors because of discretization (as discussed above) together with commutator errors that include all Hamiltonian terms, in $\hat H_0$ and $\hat V (t)$, as well as commutators of terms in $V$ at different times. 
Altogether, the Trotter error bound is
\begin{align}
    \norm{\hat U_{\rm exact} - \hat U _{\rm Trotter}} = \mathcal{O}\left( \frac{t^2}{m_T} \left( \max_{s_1,s_2}\norm{\comm{\hat H(s_1)}{\hat H(s_2)}} + \max_s \norm{\dot{ \hat{H}}(s)} \right) 
    +t \left( \frac{t}{m_T} \right)^q \max_{i,...,j}\norm{\comm{\hat H(s)^{(i)}}{\comm{...}{\hat H(s)^{(j) } } } } \right) \, ,
    \label{eq:appendix_general_Trotter_error}
\end{align}
where, in both cases, we labelled the $j^\text{th}$ term in a Hamiltonian $\hat H(t)$ as $\hat H(t)^{(j)}$. For the driven XY model trotterized in linear order, we have
\begin{align}
\label{eq:appendix_e13}
    \norm{\hat U_{\rm exact} - \hat U _{\rm QHiFFS}} &= \mathcal{O}\left( \frac{J_Z^k (J_X + J_Y) t}{\omega^{k+1}} + \frac{t^2}{m_Q} J_X J_Y \right) \\
    \norm{\hat U_{\rm exact} - \hat U _{\rm Trotter}} &= \mathcal{O}\left( \frac{t^2}{m_T} \left( J_Z \omega + J_X J_Y + 3 J_X J_Z + 3 J_Y J_Z \right) \right).
    \label{eq:appendix_e14}
\end{align}
We can read off two features. First, the Trotterization of the effective Hamiltonian in QHiFFS only requires splitting the $X \otimes X$ and $Y \otimes Y$ terms leaving a single commutator error term proportional to $J_X J_Y$ in the QHiFFS error bound, while in standard Trotterization all three combinations $J_X J_Y, J_X J_Z $ and $ J_Y J_Z$ survive. Second, the Trotter error bound has a linear contribution in $\omega$ while the QHiFFS error bound has an inverse dependence $\frac{1}{\omega^{k+1}}$. This emphasizes yet again the approximation guarantee of QHiFFS for high frequencies, and Trotterization for low frequencies and time-independent Hamiltonians. The error from time-ordering contains the estimate $\max_s \abs{\cos\left( s+\frac{t}{m} \right) - \cos(s)} \leq 2$ in the $J_X J_Z$ and $J_Y J_Z$ contributions. If a second order Trotter formula is used together with a second order discretization scheme that shifts the times at which $H(s)$ is evaluated by $\frac{\delta t}{2}$, then the corresponsing error terms scale with $\frac{t^3}{m^2}$ instead. To be precise,
\begin{align}
    \norm{\hat U_{\rm exact} - \hat U _{\rm QHiFFS}} &= 
    \mathcal{O}\left( \frac{J_Z^k (J_X + J_Y) t}{\omega^{k+1}} + \frac{t^3}{m_Q^2} (J_X^2 J_Y + J_X J_Y^2) \right) 
    \label{eq: QHiFFS_norm_XX_YY_cos_wt_ZZ}
    \\
    \norm{\hat U_{\rm exact} - \hat U _{\rm Trotter}} &= \mathcal{O}\left( 
    \frac{t^2}{m_T} (2 J_X J_Z + 2 J_Y J_Z) 
    + \right.\nonumber \\
    &\qquad \quad \left. 
    \frac{t^3}{m_T^2} \left( J_Z \omega^2 + J_X^2 J_Y + J_X J_Y^2 + J_X^2 J_Z + J_X J_Z^2 + J_Z^2 J_Y + J_Z J_Y^2 + J_X J_Y J_Z \right) 
    \right)
    \label{eq: Std_Trotter_norm_XX_YY_cos_wt_ZZ}.
\end{align}

 The quadratic contributions remain from the commutator error of the Hamiltonian with itself at different times. The improvement for high frequencies, however, becomes more pronounced as the time discretization takes into account higher derivatives that scales with $\omega^2$.

\end{document}